\documentclass[10pt]{article}
\pdfoutput=1
\usepackage{graphicx}
\usepackage{geometry}
\usepackage{amsfonts}
\usepackage{amsmath}
\usepackage{amssymb}
\usepackage{mathptmx}
\usepackage{hyperref}
\usepackage{cancel}
\usepackage{textcomp}
\usepackage{bm}
\usepackage{times}
\usepackage{epsfig}
\usepackage{xcolor}
\usepackage{colortbl}
\usepackage{slashed}
\usepackage{booktabs}
\usepackage{multirow}
\usepackage{latexsym}
\usepackage{verbatim}
\usepackage{extarrows}
\usepackage{rotating}
\definecolor{mygray}{gray}{.95}
\definecolor{intnull}{RGB}{213,229,255}
\usepackage{arydshln}
\usepackage{chngcntr}
\usepackage[title]{appendix}
\newcommand{\calL}{\mathcal{L}}
\newcommand{\calO}{\mathcal{O}}

\newcommand{\Tr}{\rm Tr}

\newcommand{\eV}{\rm eV}

\newcommand{\GeV}{\rm GeV}
\newcommand{\TeV}{\rm TeV}

\begin{document}
\baselineskip=16pt

\pagenumbering{arabic}

\vspace{1.0cm}

\begin{center}
{\Large\sf Renormalization group evolution of dimension-seven  operators in standard model effective field theory and relevant phenomenology
}
\\[10pt]
\vspace{.5 cm}

{Yi Liao~$^{a,b}$\footnote{liaoy@nankai.edu.cn} and Xiao-Dong Ma~$^{a}$\footnote{maxid@mail.nankai.edu.cn}}

{
$^a$~School of Physics, Nankai University, Tianjin 300071, China
\\
$^b$ Center for High Energy Physics, Peking University, Beijing 100871, China}

\vspace{2.0ex}

{\bf Abstract}
\end{center}

We showed in a previous publication that there are six independent dimension-seven operators violating both lepton and baryon numbers ($L=-B=1$) and twelve ones violating lepton but preserving baryon number ($L=2,~B=0$) in standard model effective field theory, and we calculated one-loop renormalization for the former six operators. In this work we continue our efforts on renormalization of the operators. It turns out this could become subtle because the operators are connected by nontrivial relations when fermion flavors are counted. This kind of relations does not appear in lower dimensional operators. We show how we can extract anomalous dimension matrix for a flavor-specified basis of operators from counterterms computed for the above flavor-blind operators without introducing singular inverse Yukawa coupling matrices. As a phenomenological application, we investigate renormalization group effects on nuclear neutrinoless double $\beta$ decay. We also discuss very briefly its analog in the meson sector, $K^\pm\to\pi^\mp\mu^\pm\mu^\pm$, and indicate potential difficulties to compute its decay width.

\newpage

\section{Introduction}

The standard model effective field theory (SMEFT) is a systematic approach to low energy effects from unknown high-scale new physics. It has become practically more and more important due to null results in searching for new particles of mass below the electroweak scale. In this approach the dynamical degrees of freedom are restricted to those in the standard model (SM) and the SM interactions are nothing but the leading ones in an infinite tower of interactions:
\begin{equation}
\mathcal{L}_\textrm{SMEFT}=\mathcal{L}_\textrm{SM}+\sum_{d\geq 5, i}C^d_i\mathcal{O}^d_i.
\label{LSMEFT}
\end{equation}
Suppressing gauge-fixing related terms the SM Lagrangian is
\begin{eqnarray}
\label{sml}
\nonumber
\mathcal{L}_\textrm{SM}&=&
-\frac{1}{4}\sum_X X_{\mu\nu}X^{\mu\nu}+(D_\mu H)^\dagger(D^\mu H)
-\lambda\left(H^\dagger H-\frac{1}{2}v^2\right)^2
\\
&&
+\sum_\Psi\bar{\Psi}i \slashed{D}\Psi
-\left[\bar{Q}Y_u u \tilde{H}+\bar{Q}Y_d d H+\bar{L}Y_e e H +\mbox{h.c.}\right],
\end{eqnarray}
where $X$ stands for the three gauge field strengths of respective couplings $g_{3,2,1}$ for the gauge group $SU(3)_C\times SU(2)_L\times U(1)_Y$, and $\Psi$ covers all fermions including left-handed quark and lepton doublets $Q,~L$ and right-handed quark and lepton singlets $u,~d,~e$ with appropriate gauge covariant derivatives $D_\mu$. $H$ is the Higgs doublet field with $\tilde H_i=\epsilon_{ij}H^*_j$. The Yukawa couplings $Y_{u,d,e}$ are generic complex $3\times 3$ matrices with three generations of fermions.

SMEFT is expected to work in the energy range between certain new physics scale $\Lambda$ and the electroweak scale set by the vacuum expectation value $v$ of the Higgs field $H$. It thus builds a bridge between unknown new physics above $\Lambda$ and physical processes explored in current experiments below $v$. New physics effects are organized in a tower of effective operators $\calO_i^d$ of increasing canonical dimension $d\ge 5$ whose impacts are measured by Wilson coefficients $C_i^d$ of decreasing relevance. These high-dimensional operators are generated by integrating out heavy degrees of freedom in new physics, and they respect SM gauge symmetries but not necessarily its accidental symmetries such as lepton and baryon number conservation. Their coefficients are generally suppressed by the new physics scale $\Lambda$, $C_i^d\sim\Lambda^{4-d}$. An important task in this endeavor is to establish a correct basis of operators in each dimension and to work out their renormalization group evolution (RGE) effects from $\Lambda$ to $v$ due to SM interactions. When a type of new physics is specified, this facilitates direct comparison of new physics with low energy measurements with the help of matching calculations at the scales $\Lambda$ and $v$.

It has been known for a long time that the dimension-five (dim-5) operator is unique~\cite{Weinberg:1979sa} and generates an effective Majorana neutrino mass. The complete and independent sets of dim-6 and dim-7 operators have been constructed in Refs.~\cite{Buchmuller:1985jz, Grzadkowski:2010es} and~\cite{Lehman:2014jma,Liao:2016hru} respectively. These operators have also been examined in an independent approach based on Hilbert series which counts fermion flavor structures, and further extended to even higher dimensions~\cite{Lehman:2015via,Henning:2015daa,
Lehman:2015coa,Henning:2015alf,Henning:2017fpj}. If the SM is generalized by sterile neutrinos of mass below the electroweak scale, there will be additional operators at each dimension, see Refs.~\cite{Aparici:2009fh,
delAguila:2008ir, Bhattacharya:2015vja,Liao:2016qyd} for discussions on operators up to dim-7 that involve sterile neutrinos. The renormalization group running of effective operators due to SM interactions is important for precision phenomenological analysis. Currently the 1-loop RGE has been completed for the dim-5~\cite{Antusch:2001ck} and dim-6~\cite{Grojean:2013kd,Elias-Miro:2013gya, Elias-Miro:2013mua, Jenkins:2013zja,Jenkins:2013wua,Alonso:2013hga,Alonso:2014zka} operators. While all dim-7 operators violate lepton number, they are classified into two subsets: one has 12 baryon number conserving operators and the other has 6 baryon number violating operators; see table \ref{tab1}. The RGE analysis for the subset that violates baryon number has been done in Ref.~\cite{Liao:2016hru}. All these 1-loop results for anomalous dimension matrices follow interesting patterns~\cite{Alonso:2014rga,Cheung:2015aba} and simple perturbative power counting rules~\cite{Jenkins:2013sda,Liao:2017amb}. The purpose of this work is to finish the 1-loop RGE analysis of dim-7 operators by completing the subset that conserves baryon number.

As we will discuss in sections \ref{sec2} and \ref{sec3} and demonstrate in table~\ref{tab2}, rich and nontrivial flavor relations among operators first appear at dimension seven. This makes analysis of dim-7 operators very different from dim-5 and dim-6 operators. While the above mentioned $12+6$ operators are convenient for extracting 1-loop counterterms when renormalizing them, they cannot be directly employed to do RGE in phenomenological analysis where specific flavors of fermions have to be discriminated because not all of them are flavor independent. For brevity we say they are in a {\em flavor-blind basis} though this is not a basis in the strict sense of the word; in contrast we call the set of genuinely independent operators taking into account flavor relations a {\em flavor-specified basis}. The anomalous dimension matrix should be defined and computed for this latter basis of operators. Since the flavor relations involve Yukawa coupling matrices whose entries are mostly small, a suitable choice of a flavor-specified basis should avoid the appearance of inverse Yukawa matrices when expressing dependent operators in terms of those in the basis. We will show an example of such choices in section \ref{sec3}. Our result for the anomalous dimension matrix still follows the patterns explained by non-renormalization theorem~\cite{Cheung:2015aba} and power counting rules~\cite{Liao:2017amb}.

Since all dim-7 operators violate lepton number, their effects are presumably small and can only be explored in high precision measurements. For the purpose of illustration we will study nuclear neutrinoless double $\beta$ decays ($0\nu\beta\beta$) and discuss very briefly lepton-number violating decays of the charged kaons $K^\pm\to\pi^\mp\ell^\pm\ell^\pm$. We will find that these processes generally involve new low-energy mechanisms for lepton number violation beyond the widely studied neutrino mass insertion. As these new mechanisms contain hadronic matrix elements that have not yet been well investigated for kaon decays in the literature, we defer an appropriate phenomenological analysis to our future work.

\begin{table}[!ht]
\centering
\begin{tabular}{|l|c|l|c|}
 \multicolumn{2}{c}{$\psi^2H^4$} &  \multicolumn{2}{c}{ $\psi^2H^3D$}
\\
\hline
$\mathcal{O}_{LH}$ & $\epsilon_{ij}\epsilon_{mn}(L^iCL^m)H^jH^n(H^\dagger H)$ & $\mathcal{O}_{LeHD}$ & $\epsilon_{ij}\epsilon_{mn}(L^iC\gamma_\mu e)H^jH^miD^\mu H^n$
\\
\hline
 \multicolumn{2}{c}{$\psi^2H^2D^2$}& \multicolumn{2}{c}{$\psi^2H^2X$}
\\
\hline
$\mathcal{O}_{LHD1}$ &$\epsilon_{ij}\epsilon_{mn}(L^iCD^\mu L^j)H^m(D_\mu H^n)$ &$\mathcal{O}_{LHB}$    &$ g_1\epsilon_{ij}\epsilon_{mn}(L^iC\sigma_{\mu\nu}L^m)H^jH^nB^{\mu\nu}$ \\
$\mathcal{O}_{LHD2}$  & $\epsilon_{im}\epsilon_{jn}(L^iCD^\mu L^j)H^m(D_\mu H^n)$ & $\mathcal{O}_{LHW}$  &$g_2\epsilon_{ij}(\epsilon \tau^I)_{mn}(L^iC\sigma_{\mu\nu}L^m)H^jH^nW^{I\mu\nu}$ \\
\hline
  \multicolumn{2}{c}{$\psi^4D$}  &   \multicolumn{2}{c}{$\psi^4H$}\\ \hline
$\mathcal{O}_{\bar{d}uLLD}$ & $\epsilon_{ij}(\bar{d}\gamma_\mu u)(L^iCiD^\mu L^j)$ & $\mathcal{O}_{\bar{e}LLLH}$ & $\epsilon_{ij}\epsilon_{mn}(\bar{e}L^i)(L^jCL^m)H^n$\\
   &  & $\mathcal{O}_{\bar{d}LQLH1}$ & $\epsilon_{ij}\epsilon_{mn}(\bar{d}L^i)(Q^jCL^m)H^n$\\
   &  & $\mathcal{O}_{\bar{d}LQLH2}$ & $\epsilon_{im}\epsilon_{jn}(\bar{d}L^i)(Q^jCL^m)H^n$\\
   &  & $\mathcal{O}_{\bar{d}LueH}$ & $\epsilon_{ij}(\bar{d}L^i)(uCe)H^j$\\
   &  & $\mathcal{O}_{\bar{Q}uLLH}$ & $\epsilon_{ij}(\bar{Q}u)(LCL^i)H^j$\\
\hline
\rowcolor{mygray}
$\mathcal{O}_{\bar{L}QddD}$ & $(\bar{L}\gamma_\mu Q)(dCiD^\mu d)$ & $\mathcal{O}_{\bar{L}dud\tilde{H}}$ & $(\bar{L}d)(uCd)\tilde{H}$\\
\rowcolor{mygray}
$\mathcal{O}_{\bar{e}dddD}$  & $(\bar{e}\gamma_\mu d)(dCiD^\mu d)$ & $\mathcal{O}_{\bar{L}dddH}$ & $(\bar{L}d)(dCd)H$\\
\rowcolor{mygray}
   &  & $\mathcal{O}_{\bar{e}Qdd\tilde{H}}$ & $\epsilon_{ij}(\bar{e}Q^{i})(dCd)\tilde{H}^j$\\
\rowcolor{mygray}
   &  & $\mathcal{O}_{\bar{L}dQQ\tilde{H}}$ & $\epsilon_{ij}(\bar{L}d)(QCQ^{i})\tilde{H}^j$ \\ \hline
\end{tabular}
\caption{Dim-7 operators in 6 classes are divided into two subsets with $L=2$ and $B=0$ and $B=-L=1$ (in gray) respectively, where
$(D_\mu H^n)$ should be understood as $(D_\mu H)^n$ etc.}
\label{tab1}
\end{table}

\section{Dimension 7 operators and their flavor structure}
\label{sec2}

For convenience we reproduce in table \ref{tab1} the $12+6$ dim-7 operators finally fixed in Ref~\cite{Liao:2016hru}. These operators are complete and independent when fermion flavors are not counted, and form the so-called flavor-blind basis introduced above. They include two subsets according to their lepton $L$ and baryon $B$  numbers, 12 operators with $L=2$ and $B=0$ and 6 ones with $B=-L=1$. They are all non-Hermitian and will be multiplied by generally complex Wilson coefficients.

\begin{table}[!ht]
\centering
\begin{tabular}{|ll|c|}
\hline
Class & Operator &  Flavor relations
\\
\hline
$\psi^2H^4$&$\mathcal{O}_{LH}$
& $\mathcal{O}^{pr}_{LH}-p\leftrightarrow r=0$
\\
\hline
$\psi^2H^3D$&$\mathcal{O}_{LeHD}$
&$\times$
\\
\hline
$\psi^2H^2D^2$&$\mathcal{O}_{LHD1}$
& $(\mathcal{O}^{pr}_{LDH1}+\mathcal{K}^{pr})-p\leftrightarrow r=0$
\\
&$\mathcal{O}_{LHD2}$
&$\Big[4\mathcal{O}^{pr}_{LHD2}+2(Y_e)_{rv}\mathcal{O}^{pv}_{LeHD}
-\mathcal{O}^{pr}_{LHW}+2\mathcal{K}^{pr}\Big]-p\leftrightarrow r=\mathcal{O}^{pr}_{LHB}$
\\
\hline
$\psi^2H^2X$
&$\mathcal{O}_{LHB}$
& $\mathcal{O}^{pr}_{LHB}+p\leftrightarrow r=0$
\\
& $\mathcal{O}_{LHW}$
& $\times$
\\
\hline
$\psi^4H$
& $\mathcal{O}_{\bar{e}LLLH}$
&$(\mathcal{O}^{prst}_{\bar{e}LLLH}+r\leftrightarrow t)-r\leftrightarrow s=0$
\\
&$\mathcal{O}_{\bar{d}LQLH1}$ &$\times$
\\
&$\mathcal{O}_{\bar{d}LQLH2}$ &$\times$
\\
& $\mathcal{O}_{\bar{d}LueH}$   &$\times$
\\
&$\mathcal{O}_{\bar{Q}uLLH}$   &$\times$
\\
\hline
$\psi^4D$
& $\mathcal{O}_{\bar{d}uLLD}$
& $\Big[\mathcal{O}^{prst}_{\bar{d}uLLD}
+(Y_d)_{vp}\mathcal{O}^{vrst}_{\bar{Q}uLLH}
-(Y_u^\dagger)_{rv}\mathcal{O}^{psvt}_{\bar{d}LQLH2}\Big]
-s\leftrightarrow t=0$
\\
\hline
\rowcolor{mygray}
$\psi^4H$
&$\mathcal{O}_{\bar{L}dud\tilde{H}}$
&$\times$
\\
\rowcolor{mygray}
&$\mathcal{O}_{\bar{L}dddH}$
& $\mathcal{O}^{prst}_{\bar{L}dddH}+s\leftrightarrow t=0, \quad \mathcal{O}^{prst}_{\bar{L}dddH}+\mathcal{O}^{pstr}_{\bar{L}dddH}
+\mathcal{O}^{ptrs}_{\bar{L}dddH}=0$
\\
\rowcolor{mygray}
&$\mathcal{O}_{\bar{e}Qdd\tilde{H}}$
&$\mathcal{O}^{prst}_{\bar{e}Qdd\tilde{H}}+s\leftrightarrow t=0$
\\
\rowcolor{mygray}
&$\mathcal{O}_{\bar{L}dQQ\tilde{H}}$
&$\times$
\\
\hline
\rowcolor{mygray}
$\psi^4D $
&$\mathcal{O}_{\bar{L}QddD}$
&$\Big[\mathcal{O}^{prst}_{\bar{L}QddD}
+(Y_u)_{rv}\mathcal{O}^{psvt}_{\bar{L}dud\tilde{H}}\Big]
-s\leftrightarrow t=-(Y_e^\dagger)_{vp}\mathcal{O}^{vrst}_{\bar{e}Qdd\tilde{H}}
-(Y_d)_{rv}\mathcal{O}^{pvst}_{\bar{L}dddH}$
\\
\rowcolor{mygray}
&$\mathcal{O}_{\bar{e}dddD}$
&$\mathcal{O}^{prst}_{\bar{e}dddD}-r\leftrightarrow s=(Y^\dagger_d)_{tv}\mathcal{O}^{pvrs}_{\bar{e}Qdd\tilde{H}}$
\\
\rowcolor{mygray}
&
&$(\mathcal{O}^{prst}_{\bar{e}dddD}+r\leftrightarrow t)-s\leftrightarrow t=(Y_e)_{vp}\mathcal{O}^{vrst}_{\bar{L}dddH}$
\\
\hline
\end{tabular}
\caption{Flavor relations for dim-7 operators. The symbol $\times$ indicates lack of such a relation.}
\label{tab2}
\end{table}

We denote fermion flavors ($p,r,s,t,...$) of an operator in the same order that fermion fields appear in the operator, and all repeated indices are implied to be summed over unless otherwise stated. For instance, $\mathcal{O}_{LH}^{pr}
=\epsilon_{ij}\epsilon_{mn}(L^i_pCL^m_r)H^jH^n(H^\dagger H)$ and  $\mathcal{O}_{\bar{d}uLLD}^{prst}
=\epsilon_{ij}(\bar{d}_p\gamma_\mu u_r)(L^i_sCiD^\mu L^j_t)$. It is easy to understand that operators involving two or three like-charge fermions may be related. With two like-charge fermions the relations are simply symmetric or antisymmetric, as is the case with the operators $\calO_{LH}^{pr}$, $\calO_{LHB}^{pr}$, and  $\calO_{\bar eQdd\tilde H}^{prst}$. With three like-charge fermions the relations generally have a mixed symmetry, and these cover the operators $\mathcal{O}_{\bar eLLLH}^{prst}$ and $\mathcal{O}_{\bar LdddH}^{prst}$. Nontrivial flavor relations exist for operators that involve at least one derivative, including $\mathcal{O}_{LHD1}^{pr}$, $\mathcal{O}_{LHD2}^{pr}$, $\mathcal{O}_{\bar duLLD}^{prst}$, $\mathcal{O}_{\bar LQddD}^{prst}$, and $\calO_{\bar edddD}^{prst}$. This explains why this feature appears first at dimension seven but not lower dimensions, and it is expected to prevail at higher dimensions. We list all flavor relations in table \ref{tab2}, in which a shortcut is used,
\begin{eqnarray}
\mathcal{K}^{pr}=(Y_u)_{vw}\mathcal{O}^{vwpr}_{\bar{Q}uLLH}
-(Y^\dagger_d)_{vw}\mathcal{O}^{vpwr}_{\bar{d}LQLH2}
-(Y^\dagger_e)_{vw}\mathcal{O}^{vwpr}_{\bar{e}LLLH}.
\end{eqnarray}
The derivation of flavor relations involves judicious applications of equations of motion (EoM) in SM, integration by parts (IBP), and Fierz identities (FI) for fermion bilinears and $SU(2)$ group generators. As an example, we derive the relation for operators $\mathcal{O}^{prst}_{\bar{d}uLLD}$ in class-$\psi^4D$:
\begin{eqnarray}
\label{flavor}
\nonumber
&&\mathcal{O}^{prst}_{\bar{d}uLLD}-s\leftrightarrow t
\\
\nonumber
&&=\epsilon_{ij}(\bar{d}_p\gamma_\mu u_r)(L^i_sCiD^\mu L^j_t)
-s\leftrightarrow t
\\
\nonumber
&&\xlongequal{\text{IBP}}-\epsilon_{ij}(\bar{d}_pi
\overleftarrow{\slashed{D}} u_r)(L^i_s C L^j_t)
-\epsilon_{ij}(\bar{d}_pi\slashed{D}u_r)(L^i_s C L^j_t)
\\
\nonumber
&&\xlongequal{\text{EoM}}(Y_d)_{vp}\Big[\epsilon_{ij}\delta_{mn}
(\bar{Q}^m_vu_r)(L^i_s C L^j_t)H^n\Big]
-(Y^\dagger_u)_{rv}\Big[\epsilon_{ij}\epsilon_{mn}
(\bar{d}_pQ^m_v)(L^i_s C L^j_t)H^n\Big]
\\
&&
\xlongequal{\text{FI}}\Big[-(Y_d)_{vp}\mathcal{O}^{vrst}_{\bar{Q}uLLH}
+(Y^\dagger_u)_{rv}\mathcal{O}^{psvt}_{\bar{d}LQLH2}\Big]
-s\leftrightarrow t,
\end{eqnarray}
where the total derivative term is neglected in the second equality, EoM's for quark fields $d$ and $u$ are implemented in the third, and finally the Fierz identities are applied to cast the operators thus obtained into the ones listed in table~\ref{tab1}.

All of the above independent flavor relations must be applied to remove redundant operators before a flavor-specified basis is achieved. We have checked that the dimension of such a basis coincides with counting of independent operators in the Hilbert series approach~\cite{Henning:2015alf}; for instance, with one (three) generation(s) of fermions there are in total 30(1542) independent operators in the basis when Hermitian conjugates of the operators are also counted. In principle the choice of a basis is arbitrary for RGE analysis~\cite{Arzt:1993gz}. In practice however, since the above nontrivial relations involve Yukawa coupling matrices whose entries are generally small, one should avoid using their inverse when removing redundant operators. Note that even if one deletes redundant operators from a basis at the start they can reappear by renormalization of chosen basis operators. It is thus important to choose a suitable basis so that no singular relations would be appealed to when recasting those redundant operators in terms of basis operators. Inspection of the relations suggests the following priority of reserving operators in the flavor-specified basis: first the operators without a derivative, then the ones with one derivative and finally the ones with two derivatives. In each step we exploit the relations to remove redundant operators. For instance, one appropriate choice would be as follows. We include the following operators in the basis: for the subset $L=2,~B=0$,
\begin{eqnarray}
&&
\frac{1}{2}\big(\mathcal{O}_{LH}^{pr}+\mathcal{O}_{LH}^{rp}\big),\quad \mathcal{O}_{LeHD}^{pr}, \quad \frac{1}{2}\big(\mathcal{O}_{LHD1}^{pr}+\mathcal{O}_{LHD1}^{rp}\big), \quad \frac{1}{2}\big(\mathcal{O}_{LHD2}^{pr}+\mathcal{O}_{LHD2}^{rp}\big), \quad \frac{1}{2}\big(\mathcal{O}_{LHB}^{pr}-\mathcal{O}_{LHB}^{rp}\big),
\nonumber
\\
&&\mathcal{O}_{LHW}^{pr}, \quad
	\mathcal{O}_{\bar{d}LQLH1}^{prst},\quad \mathcal{O}_{\bar{d}LQLH2}^{prst}, \quad \mathcal{O}_{\bar{d}LueH}^{prst},
	\quad \mathcal{O}_{\bar{Q}uLLH}^{prst}, \quad \frac{1}{2}\big(\mathcal{O}_{\bar{d}uLLD}^{prst}
+\mathcal{O}_{\bar{d}uLLD}^{prst}\big),
\nonumber
\\
&&\frac{1}{4}\big(\mathcal{O}_{\bar{e}LLLH}^{prst}+  \mathcal{O}_{\bar{e}LLLH}^{ptsr}+  \mathcal{O}_{\bar{e}LLLH}^{psrt}+  \mathcal{O}_{\bar{e}LLLH}^{ptrs} \big)\textrm{ (with at least two of $r,s,t$ being equal)},
\nonumber
\\
&&\mathcal{O}_{\bar{e}LLLH}^{prst}, ~\mathcal{O}_{\bar{e}LLLH}^{prts},~  \mathcal{O}_{\bar{e}LLLH}^{psrt}, ~\mathcal{O}_{\bar{e}LLLH}^{pstr}~ (\textrm{for }r<s<t),
\end{eqnarray}
and for the subset $B=-L=1$,
\begin{eqnarray}
&&\mathcal{O}_{\bar{L}dud\tilde{H}}^{prst}, \quad  \frac{1}{2}\big(\mathcal{O}_{\bar{e}Qdd\tilde{H}}^{prst}
-\mathcal{O}_{\bar{e}Qdd\tilde{H}}^{prts}\big), \quad \mathcal{O}_{\bar{L}dQQ\tilde{H}}^{prst},\quad \frac{1}{2}\big(\mathcal{O}_{\bar{L}QddD}^{prst}
+\mathcal{O}_{\bar{L}QddD}^{prts}\big),
\nonumber
\\
&&\frac{1}{2}\big( \mathcal{O}_{\bar{L}dddH}^{prst}
-\mathcal{O}_{\bar{L}dddH}^{prts}\big)
\textrm{ (with at least two of $r,s,t$ being equal)},
\nonumber
\\
&&\mathcal{O}_{\bar{L}dddH}^{prst}, ~  \mathcal{O}_{\bar{L}dddH}^{pstr}~(\textrm{for }r<s<t),~
\frac{1}{6}\big(\mathcal{O}_{\bar{e}dddD}^{prst}+5\text{ permutations of }(r, s, t)\big),
\end{eqnarray}
where the indices $p,r,s,t$ take values $1,2,3$ with three generations of fermions. All other operators in the flavor-blind basis are redundant ones and can be expressed by nonsingular flavor relations as a linear sum of the above operators in the flavor-specified basis.

\section{Renormalization of operators and extraction of anomalous dimension matrix}
\label{sec3}

The effective interaction involving a high-dimensional operator is typically induced at a high energy scale. When it is applied to a process or matched to an effective field theory at a low energy scale, naive perturbation theory could be spoiled by large logarithms of the ratio of the two scales. Renormalization group equation offers a systematic approach to improving perturbation theory by summing the logarithms to all orders. In doing so the correct choice of a basis of operators is a prerequisite.

We recall that we introduced two bases. The flavor-blind basis (FBB) includes all of $12+6$ operators listed in table \ref{tab1} without referring to fermion flavors. This is not a genuine basis of operators, but is very convenient for computing counterterms to  effective interactions when the latter are dressed by SM interactions. Once this is achieved, we forget about it and move on to the flavor-specified basis (FSB) to extract anomalous dimension matrix for physical applications. This is a genuine basis of operators in which all redundancy in FBB due to fermion flavor relations has been removed. But the choice of an FSB is not unique; our suggestion is that we should avoid artificial flaws such as inverse Yukawa coupling matrices when recasting redundant operators in terms of those included in the FSB. This is indeed possible according to our discussion in the last section.

Now we formulate how the above procedure is implemented. Suppose we choose an FSB (index $b$). All operators in FBB are either included in the FSB or redundant (index $r$), and they appear in the  effective Lagrangian in the form $C_b{\cal O}_b+C_r{\cal O}_r$  where $C_b,~C_r$ are Wilson coefficients and summation over $b,~r$ is implied. We stress again that the $C_r{\cal O}_r$ term is not necessary for either matching calculation or RGE and that its appearance only facilitates computing counterterms using the $12+6$ operators without specifying fermion flavors. The counterterms in $D=4-2\epsilon$ dimensions with minimal subtraction are denoted as
\begin{eqnarray}
\textrm{c.t.}=-
\big(\langle C_b\calO_b\rangle+\langle C_r\calO_r\rangle\big),
\end{eqnarray}
where $\langle C\calO\rangle$ stands for the one-loop contribution with an insertion of the effective interaction $C\calO$ dressed by SM interactions. Our results for all operators in FBB computed in $R_\xi$ gauge are listed in Appendix~\ref{app}. Once this is achieved, we only need to manipulate the $\langle C_b\calO_b\rangle$ part further to extract the anomalous dimension matrix in the chosen FSB. Noting that this part generically induces both $\calO_b$ and $\calO_r$ operators, we write it in matrix form
\begin{eqnarray}
\textrm{c.t.}=-\frac{1}{16\pi^2\epsilon}C_b^T(P\calO_b+R\calO_r)+\cdots,
\end{eqnarray}
where the dots stand for the dropped $\langle C_r\calO_r\rangle$ part and $P,~R$ are matrices appropriate for 1-column matrices $C_b,~\calO_b,~\calO_r$ which can be read off from Appendix~\ref{app}. Next we employ flavor relations to replace $\calO_r$ by $\calO_r=M\calO_b$ where $M$ is a matrix obtained from the flavor relations, so that the above counterterms become
\begin{eqnarray}
\textrm{c.t.}=-\frac{1}{16\pi^2\epsilon}C_b^T(P+RM)\calO_b+\cdots.
\end{eqnarray}
Now we define operators $\calO_b$ at scale $\mu$ and associate renormalization effects to the Wilson coefficients
\begin{eqnarray}
16\pi^2\frac{dC_b}{d\ln\mu}=\gamma C_b,
\end{eqnarray}
and the anomalous dimension matrix in the chosen FSB is computed as
\begin{eqnarray}
\gamma=-\sum_{\alpha}\rho_\alpha g_\alpha
\frac{\partial}{\partial g_\alpha}(P+RM),
\label{gm}
\end{eqnarray}
where $g_\alpha=g_{1,2,3},~Y_{e,d,u},~\lambda$, and $\rho_\alpha=2$ for $g_\alpha=\lambda$ and $\rho_\alpha=1$ otherwise.

\begin{table}
\small
\center
\setlength\tabcolsep{2pt}
\begin{tabular}{c c c c c c c c c c c c c c c c c c c c  }
\hline
& $(\omega, \bar{\omega})|\chi$  & $(5,3)|3$ & $(5,3)|3$ & $(5,3)|3$ & $(5,5)|2$ & $(5,5)|2$ & $(5,5)|2$ & $(7,3)|2$ & $(7,3)|2$ & $(7,3)|2$& $(7,3)|3$ & $(7,3)|3$ & $(7,5)|1$\\
\hline
$(\omega, \bar{\omega})|\chi$  &  $\gamma_{ij}$ & $\mathcal{O}_{LHD1}$& $\mathcal{O}_{LHD2}$& $\mathcal{O}_{\bar{d}uLLD}$ & $\mathcal{O}_{LeHD}$ & $\mathcal{O}_{\bar{d}LueH}$ & $\mathcal{O}_{\bar{Q}uLLH}$  & $\mathcal{O}_{\bar{e}LLLH}$ & $\mathcal{O}_{\bar{d}LQLH1}$ & $\mathcal{O}_{\bar{d}LQLH2}$ &  $\mathcal{O}_{LHB}$   & $\mathcal{O}_{LHW}$  & $\mathcal{O}_{LH}$
\\
\hline
$(5,3)|3$ &$\mathcal{O}_{LHD1}$&$g^2$ &$g^2$ &$g^2$& \cellcolor{mygray}0 & \cellcolor{mygray}0 &\cellcolor{mygray}0 &\cellcolor{mygray}0 &\cellcolor{mygray}0 & \cellcolor{mygray}0 & \cellcolor{mygray}0 & \cellcolor{mygray}0 &\cellcolor{mygray}0
\\
\hline
$(5,3)|3$&$\mathcal{O}_{LHD2}$&$g^2$ &$g^2$ &0 & \cellcolor{mygray}0 &\cellcolor{mygray}0 & \cellcolor{mygray}0 &\cellcolor{mygray}0 & \cellcolor{mygray}0 & \cellcolor{mygray}0 & \cellcolor{mygray}0 & \cellcolor{mygray}0 & \cellcolor{mygray}0
\\
\hline
$(5,3)|3$ &$\mathcal{O}_{\bar{d}uLLD}$&$g^2$&$g^2$&$g^2$&\cellcolor{mygray}0 &\cellcolor{mygray}0 &\cellcolor{mygray}0 &\cellcolor{mygray}0 &\cellcolor{mygray}0 &\cellcolor{mygray}0 &\cellcolor{mygray}0 &\cellcolor{mygray}0 &\cellcolor{mygray}0
\\
\hline
 $(5,5)|2$ & $\mathcal{O}_{LeHD}$ &$g^3$&$g^3$& 0 &$g^2$& $g^2$ & 0 &\cellcolor{mygray}0 & \cellcolor{mygray}0 &\cellcolor{mygray}0 &\cellcolor{mygray}$\sum\rightarrow0$ & \cellcolor{mygray}$\sum\rightarrow0$ &\cellcolor{mygray}0
\\
\hline
$(5,5)|2$ &$\mathcal{O}_{\bar{d}LueH}$& $g^3$ &$g^3$&$g^3$ &$g^2$&$g^2$&$g^2$ &\cellcolor{mygray}0 &\cellcolor{mygray}$Y_uY_e$&\cellcolor{mygray}$Y_uY_e$&\cellcolor{mygray}0 &\cellcolor{mygray}0 &  \cellcolor{mygray}0
\\
\hline
$(5,5)|2$ &$\mathcal{O}_{\bar{Q}uLLH}$ &$g^3$&$g^3$&$g^3$& 0 &$g^2$& $g^2$ &\cellcolor{mygray}$Y_uY_e$&\cellcolor{mygray}$Y_uY_d$&
\cellcolor{mygray}$Y_uY_d$& \cellcolor{mygray}0 &\cellcolor{mygray}0 &\cellcolor{mygray}0
\\
\hline
$(7,3)|2$ &$\mathcal{O}_{\bar{e}LLLH}$ &$g^3$&$g^3$ & 0  & \cellcolor{mygray}0 & \cellcolor{mygray}0 & \cellcolor{mygray}$Y^\dagger_uY^\dagger_e$ &$g^2$&$g^2$ &$g^2$&$g^3$ &$g^3$ & \cellcolor{mygray}0
\\
\hline
$(7,3)|2$ &$\mathcal{O}_{\bar{d}LQLH1}$   &$g^3$&$g^3$&$g^3$& \cellcolor{mygray}0& \cellcolor{mygray}$Y^\dagger_uY^\dagger_e $ & \cellcolor{mygray}$Y^\dagger_uY^\dagger_d$ &$g^2$&$g^2$&$g^2$& $g^3$&$g^3$& \cellcolor{mygray}0
\\
\hline
$(7,3)|2$ &$\mathcal{O}_{\bar{d}LQLH2}$   &$g^3$&$g^3$&$g^3$&\cellcolor{mygray}0 &\cellcolor{mygray}$Y^\dagger_uY^\dagger_e $&\cellcolor{mygray}$Y^\dagger_uY^\dagger_d$  &$g^2$&$g^2$&$g^2$& $g^3$& $g^3$ &\cellcolor{mygray}0
\\
\hline
$(7,3)|3$ & $\mathcal{O}_{LHB}$ &$g^2$ &$g^2$ & 0 &\cellcolor{mygray}$Y^\dagger_e$ &\cellcolor{mygray}0 &\cellcolor{mygray}0 &$g^1$&$g^1$&0 &$g^2$&$g^2$&\cellcolor{mygray}0
\\
\hline
 $(7,3)|3$ &$\mathcal{O}_{LHW}$   &$g^2$&$g^2$ &  0  & \cellcolor{mygray}$Y^\dagger_e$&\cellcolor{mygray}0 &\cellcolor{mygray}0 &$g^1$&$g^1$&$g^1$&$g^2$&$g^2$ &\cellcolor{mygray}0
\\
\hline
$(7,5)|1$ &$\mathcal{O}_{LH}$   &$g^4$&$g^4$& 0  &$g^3$& 0 &$g^3$&$g^3$&$g^3$&0& 0 &  $g^4$  &$g^2$
\\
\hline
\end{tabular}
\caption{The structure and perturbative power counting of the anomalous dimension matrix $\gamma_{ij}$ for the subset of dim-7 operators with $L=2,~B=0$. Also shown are holomorphic ($\omega$) and antiholomorphic ($\bar\omega$) weights and perturbative power counting ($\chi$) of the operators. The entries with $\sum\to 0$ indicate that Yukawa coupling terms happen to cancel each other.}
\label{tab3}
\end{table}

We conclude this section with a brief discussion of the structure in the anomalous dimension matrix. The structure at one loop can be understood in terms of a nonrenormalization theorem~\cite{Cheung:2015aba} and perturbative power counting rules~\cite{Liao:2017amb}. For the subset of dim-7 operators with $B=-L=1$ this was studied in detail in Ref.~\cite{Liao:2017amb}, and we thus concentrate on the other subset with $L=2,~B=0$ whose result is shown in table~\ref{tab3}. According to Ref.~\cite{Cheung:2015aba} each operator $\calO$ is assigned with a holomorphic weight $\omega(\calO)$ and an antiholomorphic weight $\bar\omega(\calO)$, and the nonrenormalization theorem asserts that up to nonholomorphic Yukawa couplings an operator $\calO_i$ can only be renormalized by an operator $\calO_j$ if $\omega(\calO_i)\ge\omega(\calO_j)$ and $\bar\omega(\calO_i)\ge\bar\omega(\calO_j)$ are both true, or to put it in another way, $\gamma_{ij}=0$ when $\omega(\calO_i)<\omega(\calO_j)$ or $\bar\omega(\calO_i)<\bar\omega(\calO_j)$. This explains the zeros in gray up to Yukawa couplings in table~\ref{tab3}. The other zeros in the table reflect the simple fact that there happens to be no one-loop diagrams. We stress that flavor relations shown in tabel~\ref{tab2} do not spoil the nonrenormalization theorem. For simple relations without involving Yukawa couplings this is obvious. For nontrivial relations involving Yukawa couplings which are brought about by EoMs, there is no theorem at all in the first place. The perturbative orders of the remaining entries in $\gamma$ can be determined by power counting rules~\cite{Liao:2017amb}. The basic idea is that all terms in the SM Lagrangian $\calL_\textrm{SM}$ are treated as same order in perturbation theory. This fixes the perturbative order of each building block and thus that of each operator, $\chi[\calO_i]$, up to a common additive constant. The perturbative order of $\gamma$ is then determined at one loop to be $\chi[\gamma_{ij}]=2+\chi[\mathcal{O}_j]-\chi[\mathcal{O}_i]$. In this counting we treat all of the couplings $g_{1,2,3},~Y_{e,u,d},~\sqrt{\lambda}$ as the same order $g$. We note again that flavor relations are automatically consistent with power counting since none of manipulations in establishing the relations would change perturbative order of an operator.

\section{Phenomenology of dim-7 operators}
\label{sec4}

We studied in a previous work~\cite{Liao:2016hru} the proton decay $p\to\pi^+\nu$ induced by the subset of dim-7 operators with $B=-L=\pm 1$. In this section we study some phenomenology of the other subset of operators with $L=\pm 2$ and $B=0$. We will improve a previous analysis~\cite{Cirigliano:2017djv,Cirigliano:2018yza} on nuclear neutrinoless double $\beta$ decay by incorporating complete one-loop SM RGE effects from a high scale to the electroweak scale. We will also discuss briefly its counterpart in the meson sector, i.e., the rare decays $K^\pm\to\pi^\mp\ell^\pm\ell^\pm$, which have been severely constrained for the muon-pair final state to be $\textrm{Br}(K^{\pm}\rightarrow\mu^{\pm}\mu^{\pm}\pi^{\mp})< 8.6\times10^{-11}~(90\%\textrm{ CL})$~\cite{CERNNA48/2:2016tdo}.

\begin{figure}
\centering
\includegraphics[width=12cm]{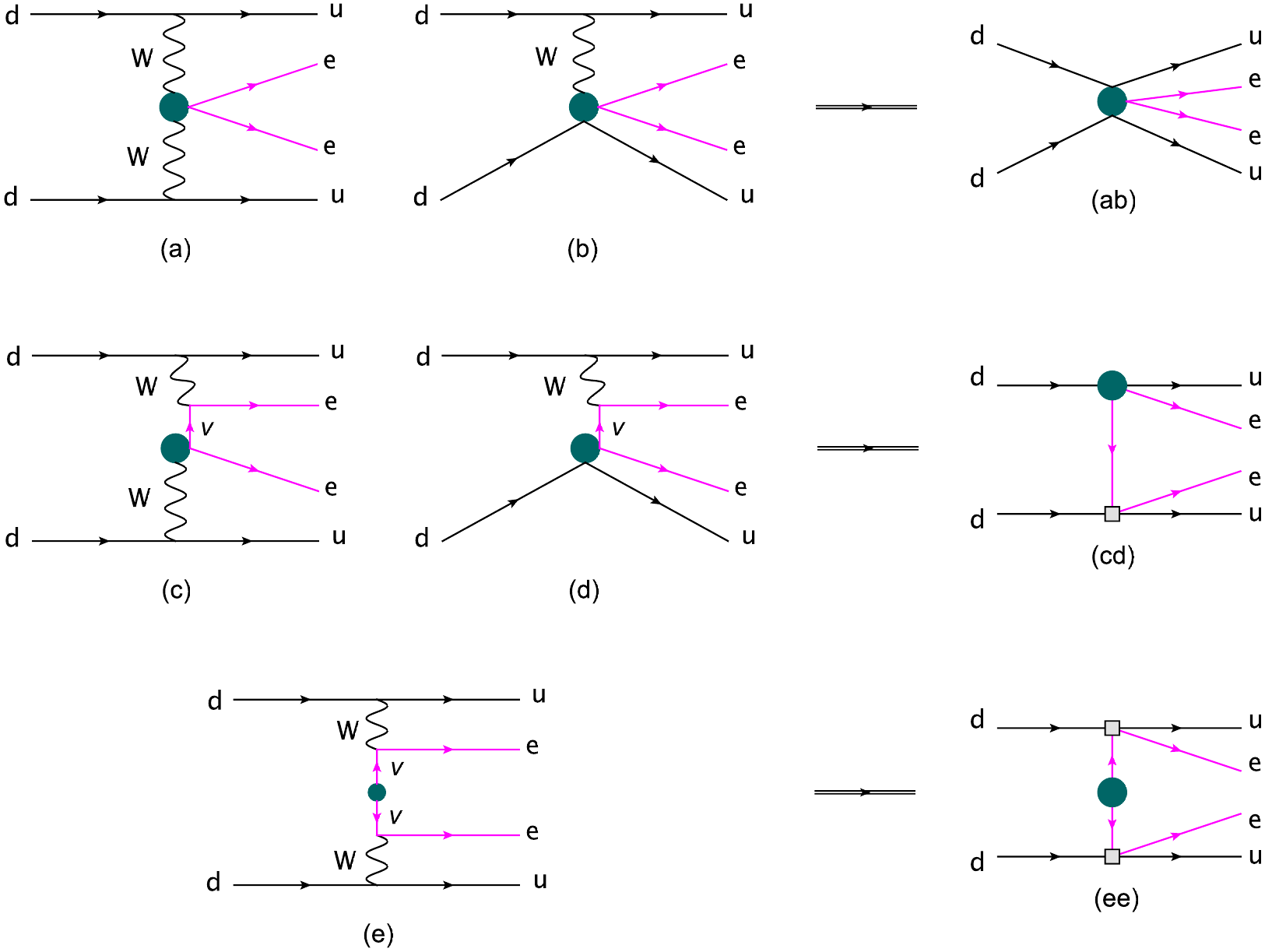}
\caption{Quark-level Feynman diagram for $0\nu\beta\beta$ from dim-7 (and dim-5 in (e)) operators (heavy dots) in SMEFT.}
\label{fig1}
\end{figure}

Nuclear neutrinoless double $\beta$ decay ($0\nu\beta\beta$) has been so far the most extensively studied process searching for lepton number violation; see, e.g., Ref.~\cite{Rodejohann:2011mu} for a review. Attributing its source to a mechanism responsible for light Majorana neutrino mass, the current experimental limit on the process translates to a bound on the effective neutrino mass  $m_{\beta\beta}<0.1~\eV$~\cite{KamLAND-Zen:2016pfg,Agostini:2018tnm}, and this bound is expected to be pushed down further to $m_{\beta\beta}<0.015~\eV$~\cite{Kharusi:2018eqi} in the near future. In the framework of SMEFT there are additional mechanisms that are not directly related to the light neutrino masses as shown in Fig.~\ref{fig1}. The heavy dot in the figure represents dim-7 effective interactions studied in previous sections (and also the dim-5 effective mass operator in subgraph (e)) that are obtained from the 12 operators in table~\ref{tab1} upon sending $H$ to its vacuum expectation value $v/\sqrt{2}$. As we go down further to lower energy scale at which the weak gauge bosons are integrated out, the diagrams in the left panel will be matched to those in the right where the box stands for the SM four-Fermi weak interactions. It is clear that there are three classes of contributions: short-range (or contact) interaction, long-range interaction, and light neutrino mass insertion, which we will study one by one below. To simplify the matter a bit, we assume that all quark and lepton mixing matrix elements have already been incorporated in the Wilson coefficients.

The short-range interaction amounts to the following dim-9 operators which are generated from dim-7 operators $\calO^{pr\dagger}_{LHD1}$, $\calO^{pr\dagger}_{LHW}$ and $\calO^{prst\dagger}_{\bar{d}uLLD}$ and the SM four-Fermi weak interactions:
\begin{eqnarray}
\mathcal{L}_\textrm{S}=(\overline{u}\gamma^\mu P_Ld)\big[
C_{\textrm{S}1} (\overline{u}\gamma_\mu P_Ld)
+C_{\textrm{S}2}(\overline{u}\gamma_\mu P_Rd)
\big](\overline{e}P_Re^C),
\end{eqnarray}
where
\begin{eqnarray}\label{eq13}
(C_{\textrm{S}1},C_{\textrm{S}2})=-2\sqrt{2}G_F
\big(C_{LHD1}^{ 11\dagger}+4C_{LHW}^{11\dagger },C_{\overline{d}uLLD}^{1111\dagger}\big),
\end{eqnarray}
and $G_F$ is the Fermi constant. Neutrinoless double $\beta$ decay has been studied in EFT below the weak scale in Refs.~\cite{Pas:2000vn,Deppisch:2012nb, Horoi:2017gmj}. Relations to the $\epsilon$ parameters in Ref.~\cite{Horoi:2017gmj} are
\begin{eqnarray}
\label{c1}
\big(C_{LHD1}^{11\dagger }+4C_{LHW}^{11\dagger },C_{\overline{d}uLLD}^{1111\dagger}\big)
=-\frac{\sqrt{2}G_F}{m_p}\big(\epsilon_3^{LLR},\epsilon_3^{LRR}\big),
\end{eqnarray}
where $m_p$ is the proton mass. The constraints on the $\epsilon$ parameters at the proton mass scale from experiments using elements $^{48}\text{Ca},^{76}\text{Ge}, ^{82}\text{Se},^{130}\text{Te},^{136}\text{Xe}$ were worked out in Ref.~\cite{Horoi:2017gmj}, and those relevant to our analysis are reproduced in table~\ref{tab4}. As a matter of fact, translating experimental limits on half lives to those on the $\epsilon$ or $C$ parameters defined at $1\sim 2~\GeV$ is afflicted with hadronic and nuclear level uncertainties which we cannot improve in this work. These uncertainties can be order one according to the most recent estimates in Refs~\cite{Horoi:2017gmj,Cirigliano:2018yza}. We refer the interested reader to Ref~\cite{Cirigliano:2018yza} and references cited therein for a comprehensive account of the issue.

\begin{table}
	\centering
	\begin{tabular}{|c|c|c|c|c|c|}
		\hline
		& $^{48}\text{Ca}$ & $^{76}\text{Ge}$ & $^{82}\text{Se}$ & $^{130}\text{Te}$ & $^{136}\text{Xe}$
		\\
		\hline
		$|\epsilon_3^{LLR}|$ & $3.8\times 10^{-7}$ & $8.9\times 10^{-9}$ & $6.7\times 10^{-8}$ & $2.0\times 10^{-8}$ & $4.1\times 10^{-9}$
		\\
		\hline
		$|\epsilon_3^{LRR}|$ & $6.3\times 10^{-7}$ & $1.4\times 10^{-8}$ & $1.1\times 10^{-7}$ & $3.2\times 10^{-8}$ & $6.7\times 10^{-9}$
		\\
		\hline
		$|\epsilon^{V+A}_{V-A}|$ & $1.1\times 10^{-7}$ & $2.2\times 10^{-9}$ & $1.7\times 10^{-8}$ & $5.1\times 10^{-9}$ & $1.1\times 10^{-9}$
		\\
		\hline
		$|\epsilon^{V+A}_{V+A}|$ & $1.3\times 10^{-5}$ & $4.3\times 10^{-7}$ & $2.2\times 10^{-6}$ & $9.3\times 10^{-7}$ & $2.0\times 10^{-7}$
		\\
		\hline
		$|\epsilon^{S+P}_{S\pm P}|$ & $3.4\times 10^{-7}$ & $7.9\times 10^{-9}$ & $6.1\times 10^{-8}$ & $1.4\times 10^{-8}$ & $2.9\times 10^{-9}$
		\\
		\hline
		$|\epsilon^{TR}_{TR}|$ & $1.8\times 10^{-8}$ & $7.9\times 10^{-10}$ & $5.9\times 10^{-9}$ & $2.0\times 10^{-9}$ & $4.2\times 10^{-10}$
		\\
		\hline
	\end{tabular}
	\caption{Upper bounds on some $|\epsilon|$ at the proton mass scale $\mu\approx m_p$ derived for various nuclei and assuming one operator is active at a time. Reproduced from Ref.~\cite{Horoi:2017gmj}.}
	\label{tab4}
\end{table}

The long-range interaction is mediated by a neutrino propagator connecting the SM four-Fermi weak interaction and the effective interactions induced by the dim-7 operators $\mathcal{O}^{pr\dagger}_{LeHD}$, $\mathcal{O}^{prst\dagger}_{\bar{d}LueH}$, $\mathcal{O}^{prst\dagger}_{\bar{d}LQLH1}$, $\mathcal{O}^{prst\dagger}_{\bar{d}LQLH2}$ and $\mathcal{O}^{prst\dagger}_{\bar{Q}uLLH}$:
\begin{eqnarray}\label{longl}
\mathcal{L}_\textrm{L}=\sum_{n=0}^{5}C_{\textrm{L}n}\mathcal{O}_{n},
\end{eqnarray}
where the SM effective interaction is
\begin{eqnarray}
\mathcal{O}_{0}
=(\overline{u}\gamma^\mu P_L d)(\overline{e}\gamma_\mu P_L \nu),~
C_{\textrm{L}0}=-2\sqrt{2}G_F,
\end{eqnarray}
and the new ones are
\begin{eqnarray}
\nonumber
\mathcal{O}_{1}&=&(\overline{u}\gamma^\mu P_L d)(\overline{e}\gamma_\mu P_R \nu^C),
\\
\nonumber
\mathcal{O}_{2}&=&(\overline{u}\gamma^\mu P_R d)(\overline{e}\gamma_\mu P_R \nu^C),
\\
\nonumber
\mathcal{O}_{3}&=&(\overline{u} P_L d)(\overline{e} P_R \nu^C),
\\
\nonumber
\mathcal{O}_{4}&=&(\overline{u}P_R d)(\overline{e}P_R \nu^C),
\\
\mathcal{O}_{5}&=&(\overline{u}\sigma^{\mu\nu} P_R d)(\overline{e}\sigma_{\mu\nu}P_R \nu^C),
\end{eqnarray}
with coefficients
\begin{eqnarray}
\nonumber
\big(C_{\textrm{L}1},C_{\textrm{L}3}\big)
&=&\frac{\sqrt{2}v}{2}\Big(-C^{11\dagger}_{LeHD},
C^{1111\dagger}_{\bar{Q}uLLH}\Big),
\\
\nonumber
\big(C_{\textrm{L}2},C_{\textrm{L}4}\big)&=&
\frac{\sqrt{2}v}{4}\Big(C^{1111\dagger}_{\bar{d}LueH},
C^{1111\dagger}_{\bar{d}LQLH1}\Big),
\\
C_{\textrm{L}5}&=&\frac{\sqrt{2}v}{16}\Big[ C^{1111\dagger}_{\bar{d}LQLH1}+2C^{1111\dagger}_{\bar{d}LQLH2}\Big].
\end{eqnarray}
Note that Fierz identities have been employed to reach the above form. Again, relations to the $\epsilon$ parameters in~\cite{Horoi:2017gmj} are
\begin{eqnarray}
\nonumber\label{c2}
\Big(C^{11\dagger}_{LeHD},C^{1111\dagger}_{\bar{Q}uLLH}\Big)&=&
\frac{4G_F}{v}\Big(-\epsilon_{V-A}^{V+A},\epsilon_{S-P}^{S+P}\Big),
\\
\nonumber
\Big(C^{1111\dagger}_{\bar{d}LueH},C^{1111\dagger}_{\bar{d}LQLH1}\Big)&=&
\frac{8G_F}{v}\Big(\epsilon_{V+A}^{V+A},\epsilon_{S+P}^{S+P}\Big),
\\
C^{1111\dagger}_{\bar{d}LQLH2}&=&
\frac{4G_F}{v}\Big[4\epsilon_{TR}^{TR}-\epsilon_{S+P}^{S+P}\Big],
\end{eqnarray}
and upper bounds on their magnitudes are also reproduced in table \ref{tab4}.

Finally the decay may be induced by insertion of a light Majorana neutrino mass in the neutrino propagator that transmits lepton number violation:
\begin{eqnarray}
\label{massterm}
\mathcal{L}_\textrm{M}
=-\frac{1}{2}v^2(C_{LH5}^{pr}+v^2C_{LH}^{pr})
(\overline{\nu_p}P_R \nu_r^C)+\text{h.c.},
\end{eqnarray}
where $C_{LH5}^{pr}$ is the Wilson coefficient of the dim-5 Weinberg operator. Since this mechanism has been extensively studied in the literature, we will concentrate on the other two. From naive dimensional analysis they are important only when the dim-5 Weinberg operator is suppressed for one reason or another.

\begin{table}
\centering
\begin{tabular}{|c|c|c|c|c|c|}
\hline
$(100~\TeV)^{-3}$& $^{48}\text{Ca}$ & $^{76}\text{Ge}$ & $^{82}\text{Se}$ & $^{130}\text{Te}$ & $^{136}\text{Xe}$
\\
\hline
$|C_{LHD1}^{11\dagger }|$ & $4.124\times10^3$ & $0.097\times10^3$ & $0.727\times10^3$ & $0.217\times10^3$ & $0.046\times10^3$
\\
\hline
$|C_{\overline{d}uLLD}^{1111\dagger}|$ & $12.36\times10^3$ & $0.274\times10^3$ & $2.149\times10^3$ & $0.625\times10^3$ & $0.131\times10^3$
\\
\hline
$|C^{11\dagger}_{LeHD}|$ & $0.021\times10^3$ & $0.4$ & $3.2$ & $1.0$ & $0.2$
\\
\hline
$|C^{1111\dagger}_{\bar{d}LueH}|$ & $4.927\times10^3$ & $0.163\times10^3$ & $0.834\times10^3$ & $0.352\times10^3$ & $0.076\times10^3$
\\
\hline
$|C^{1111\dagger}_{\bar{Q}uLLH}|$ & $0.043\times10^3$ & $1.0$ & $0.008\times10^3$ & $1.8$ & $0.4$
\\
\hline
$|C^{1111\dagger}_{\bar{d}LQLH1}|$ & $0.086\times10^3$ & $2.0$ & $0.015\times10^3$ & $3.5$ & $0.7$
\\
\hline
$|C^{1111\dagger}_{\bar{d}LQLH2}|$ & $0.068\times10^3$ & $1.3$ & $9.9$ & $1.7$ & $0.3$
\\
\hline
$|C_{LHW}^{11\dagger }|$ & $1.031\times10^3$ & $0.024\times10^3$ & $0.182\times10^3$ & $0.054\times10^3$ & $0.012\times10^3$
\\
\hline
\end{tabular}
\caption{Upper bounds on Wilson coefficients of dim-7 operators at the weak scale $\mu\approx m_W$ using RGE formulas in Ref.~\cite{Cirigliano:2017djv} and table~\ref{tab4} as initial values~\cite{Horoi:2017gmj}.}
\label{tab5}
\end{table}

Now we evolve the above bounds at the proton mass scale $\mu\approx m_p$ to those at the electroweak scale $\mu\approx m_W$ using RGE formulas in Ref.~\cite{Cirigliano:2017djv}. We adopt the physical constants recommended by the Particle Data Group~\cite{Tanabashi:2018pdg}; for instance, $m_p=0.938~\GeV$, $G_F=1.166\times10^{-5}~\GeV^{-2}$, and $v=246.22~\GeV$. The results are shown in table~\ref{tab5}. As we can see from the table, data from the nucleus $^{136}\text{Xe}$ sets the most severe constraints for all Wilson coefficients under consideration. This offers the starting point for our further RGE analysis to a higher energy scale of new physics.

To evolve from the electroweak scale to a higher scale at which dim-7 operators are generated, we first derive RGE equations relevant to $0\nu\beta\beta$ decay using eqs.~\eqref{gm} and \eqref{o11l}-\eqref{o121l}:
\begin{eqnarray}
\label{RGE1}
\nonumber
\frac{d}{d\ln\mu}C_{LHD1}^{11\dagger}&=&
\frac{1}{4\pi}\Big(-\frac{9}{10}\alpha_1
+\frac{11}{2}\alpha_2+6\alpha_t\Big)C^{11\dagger}_{LHD1}
+\frac{1}{4\pi}\Big(-\frac{33}{20}\alpha_1-\frac{19}{4} \alpha_2-2\alpha_\lambda \Big)C^{11\dagger}_{LHD2},
\\
\nonumber
\frac{d}{d\ln\mu}C_{\bar{d}uLLD}^{1111\dagger}&=&
\frac{1}{4\pi}\Big(\frac{1}{10}\alpha_1
-\frac{1}{2}\alpha_2\Big)C_{\bar{d}uLLD}^{1111\dagger},
\\
\nonumber
\frac{d}{d\ln\mu}C_{LeHD}^{11\dagger}&=&
\frac{1}{4\pi}\Big(-\frac{9}{10}\alpha_1
+6\alpha_\lambda+9\alpha_t\Big)C^{11\dagger}_{LeHD},
\\
\nonumber
\frac{d}{d\ln\mu}C_{\bar{d}LueH}^{1111\dagger}&=&
\frac{1}{4\pi}\Big(-\frac{69}{20}\alpha_1
-\frac{9}{4}\alpha_2+3\alpha_t\Big)C^{1111\dagger}_{\bar{d}LueH},
\\
\nonumber
\frac{d}{d\ln\mu}C_{\bar{Q}uLLH}^{1111\dagger}&=&
\frac{1}{4\pi}\Big(\frac{1}{20}\alpha_1
-\frac{3}{4}\alpha_2-8\alpha_3+3\alpha_t\Big)C^{1111\dagger}_{\bar{Q}uLLH},
\\
\nonumber
\frac{d}{d\ln\mu}C_{\bar{d}LQLH1}^{1111\dagger}&=&
\frac{1}{4\pi}\Big(\frac{13}{20}\alpha_1
+\frac{9}{4}\alpha_2-8\alpha_3+3\alpha_t\Big)C^{1111\dagger}_{\bar{d}LQLH1}
+\frac{1}{4\pi}\Big(6\alpha_2\Big)C^{1111\dagger}_{\bar{d}LQLH2},
\\
\nonumber
\frac{d}{d\ln\mu}C_{\bar{d}LQLH2}^{1111\dagger}&=&
\frac{1}{4\pi}\Big(-\frac{121}{60}\alpha_1
-\frac{15}{4}\alpha_2+\frac{8}{3}\alpha_3
+3\alpha_t\Big)C^{1111\dagger}_{\bar{d}LQLH2}
+\frac{1}{4\pi}\Big(-\frac{4}{3}\alpha_1
+\frac{16}{3}\alpha_3\Big)C^{1111\dagger}_{\bar{d}LQLH1},
\\
\nonumber
\frac{d}{d\ln\mu}C_{LHD2}^{11\dagger}&=&
\frac{1}{4\pi}\Big(\frac{12}{5}\alpha_1
+3\alpha_2+4\alpha_\lambda+6\alpha_t\Big)C^{11\dagger}_{LHD2}
+\frac{1}{4\pi}\Big(-8\alpha_2\Big)C^{11\dagger}_{LHD1},
\\
\frac{d}{d\ln\mu}C_{LHW}^{11\dagger}&=&
\frac{1}{4\pi}\Big(-\frac{6}{5}\alpha_1
+\frac{13}{2}\alpha_2+4\alpha_\lambda+6\alpha_t\Big)C^{11\dagger}_{LHW}
+\frac{1}{4\pi}\Big(\frac{5}{8}\alpha_2\Big)C^{11\dagger}_{LHD1}+
\frac{1}{4\pi}\Big(-\frac{9}{80}\alpha_1+\frac{11}{16}\alpha_2\Big)C^{11\dagger}_{LHD2}.
\end{eqnarray}
Note that we have kept the couplings $g_{1,2,3},~\lambda$ and the dominant top Yukawa coupling $y_t\equiv (Y_u)_{33}$ in the above equations and switched to the grand unification convention for the $g_1$ coupling, i.e., $g_1\to g_1\sqrt{3/5}$. Our $\alpha_i$ convention is standard
\begin{eqnarray}
\alpha_{1,2,3}=\frac{g_{1,2,3}^2}{4\pi},
\quad\alpha_\lambda=\frac{\lambda}{4\pi},
\quad \alpha_t=\frac{y_t^2}{4\pi},
\end{eqnarray}
which satisfy the RGE equations at one loop (for a clear exposition see Ref.~\cite{Mihaila:2012pz}):
\begin{eqnarray}
\nonumber
\frac{d\alpha_1}{d\ln\mu}&=&
\frac{1}{2\pi}\Big(\frac{1}{10}+\frac{4}{3}n_G\Big)\alpha_1^2,
\\\nonumber
\frac{d\alpha_2}{d\ln\mu}&=&
\frac{1}{2\pi}\Big(-\frac{43}{6}+\frac{4}{3}n_G\Big)\alpha_2^2,
\\\nonumber
\frac{d\alpha_3}{d\ln\mu}&=&
\frac{1}{2\pi}\Big(-11+\frac{4}{3}n_G\Big)\alpha_3^2,
\\\nonumber
\frac{d\alpha_t}{d\ln\mu}&=&
\frac{1}{2\pi}\Big(-\frac{17}{20}\alpha_1-\frac{9}{4}\alpha_2
-8\alpha_3+\frac{9}{2}\alpha_t\Big)\alpha_t,
\\
\frac{d\alpha_\lambda}{d\ln\mu}&=&
\frac{1}{4\pi}\Big(-\frac{9}{5}\alpha_1-9\alpha_2+12\alpha_t
+24\alpha_\lambda\Big)\alpha_\lambda
+ \frac{1}{8\pi}\Big(\frac{27}{100}\alpha_1^2
+\frac{9}{10}\alpha_1\alpha_2+\frac{9}{4}\alpha_2^2\Big),
\label{alpha_rge}
\end{eqnarray}
where $n_G$ is the number of fermion generations. We adopt the $\overline{\text{MS}}$ values of $\alpha_i$ at the $Z$-pole $m_Z$~\cite{Mihaila:2012pz} as our initial values
\begin{eqnarray}
\nonumber
\label{input}
&\alpha_1(m_Z)=0.0169225\pm0.0000039, \quad
\alpha_2(m_Z)=0.033735\pm0.00020, \quad
\alpha_3(m_Z)=0.1173\pm0.00069,&
\\&
\alpha_t(m_Z)=0.07514, \quad
\alpha_\lambda=0.13/4\pi, &
\end{eqnarray}
where $\alpha_\lambda$ is calculated by $4\pi\alpha_\lambda=m_H^2/(2v)$.

\begin{figure}
	\centering
	\includegraphics[width=16cm]{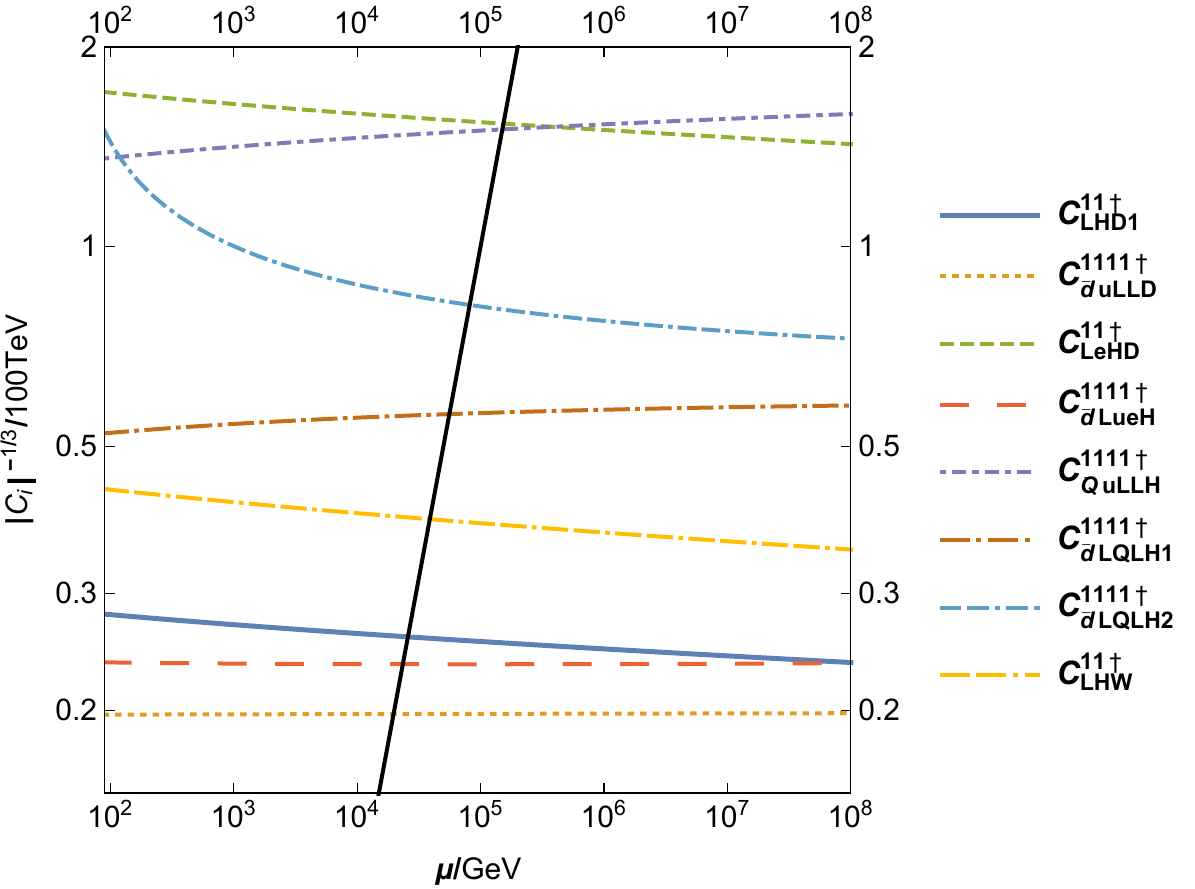}
	\caption{RGE of the Wilson coefficients relevant to $0\nu\beta\beta$ decay. The black solid line indicates $\mu=|C_i|^{-1/3}$, and roughly speaking SMEFT applies to its left region.
	}
	\label{fig2}
\end{figure}

Now we solve our RGE equations \eqref{RGE1} and \eqref{alpha_rge} numerically using the last column of table~\ref{tab5} and eq.~\eqref{input} as initial conditions. Since the operator $\calO_{LHD2}^{11\dagger}$ does not appear in the Feynman diagrams for the decay, we assume $C_{LHD2}^{11\dagger}(m_Z)=0$. Our results for the seven Wilson coefficients are shown in Fig.~\ref{fig2}. As we can see from the figure that the running effect from $100~\GeV$ to about $100~\TeV$ is mild for $C_{\bar duLLD}^{1111\dagger}$ and $C_{\bar dLueH}^{1111\dagger}$ but significant for other coefficients, and the lower limit on new  physics scale estimated naively from $|C_i|^{-1/3}$ is around $100~\TeV$ depending on the operator under consideration. Our result improves the analysis in Ref.~\cite{Cirigliano:2017djv} where only QCD interactions were considered in RGE of dim-7 operators in SMEFT, while both results agree in the order of magnitude.

Now we discuss briefly the rare decay $K^+\to\mu^+\mu^+\pi^-$ which can be considered an analog of the nuclear $0\nu\beta\beta$ decay in the meson sector. The Feynman diagrams at the quark level are also classified into three classes: short-range, long-range interactions and insertion of light Majorana neutrino masses. Since a pair of quark currents are involved, the hard core problem is to evaluate their matrix elements between the initial and final meson states of opposite charge. In quark-level Feynman diagrams the two mesons can be formed in two different manners according to whether the $W^\pm$ gauge bosons are exchanged in the $s$ or $t$ channel. In the classes of long-range interaction and mass insertion the pair of quark charged currents are defined at different points that are connected by a light neutrino propagator. This pseudoscalar level problem should be less difficult to cope with than the hadronic problem in $0\nu\beta\beta$ decay which involves nucleons as well. We note that a similar process $\pi^-\pi^-\to ee$ entering $0\nu\beta\beta$  decay has recently been worked out by lattice methods, that is due to a short-range interaction~\cite{Nicholson:2018mwc} or a long-range interaction by a neutrino propagator~\cite{Feng:2018pdq}. This result could be helpful for the evaluation of the rare $K^+$ decay by flavor $SU(3)$ symmetry. We would like to reserve for our future efforts the phenomenological analysis of the decay and related processes for the $D$ and $B$ mesons.

\section{Conclusion}
\label{sec5}

We have studied systematically the fermion flavor relations of the $(12+6)$ dim-7 operators of different Lorentz structures and field contents in SMEFT. These operators would be complete and independent without counting flavors. Some nontrivial types of flavor relations first appear at dimension seven and involve Yukawa coupling matrices. In phenomenological analysis it is necessary to choose a genuine basis for operators which must take into account individual flavor degrees of freedom. While in principle one can choose any basis, an improper choice however may bring about artefact such as inverse Yukawa coupling matrices that are almost singular in SM. We suggest a recipe to choose a proper basis: reserve priority to operators with less derivatives and along the way remove redundant operators by flavor relations.

Then we discussed how to renormalize these operators that are constrained by flavor relations. The issue is that while the $(12+6)$ operators without counting flavors are easier to work with when computing their counterterms `blindly', anomalous dimension can only be defined consistently for a set of complete and independent operators. We formulated how to form the anomalous dimension matrix for the latter from counterterms computed for the former, and listed counterterms in Appendix \ref{app}. Our one-loop results follow the patterns heralded by nonrenormalization theorem and perturbative power counting rules. As a first phenomenological application we studied renormalization group effects on nuclear neutrinoless double $\beta$ decay from the electroweak scale to certain high scale at which dim-7 operators are generated. Requiring the inverse cubic root of the Wilson coefficients to be no lower than the high scale, the running effects can still be significant for some operators. And the current experimental bound on the decay implies the inverse cubic root of the Wilson coefficients to be larger than about $100~\TeV$. We also very briefly touched upon the lepton-number violating decay $K^\pm\to\pi^\mp\mu^\pm\mu^\pm$ and pointed out its potential difficulties. We wish to come back to this process in the near future.

\vspace{0.5cm}
\noindent %
\section*{Acknowledgement}

This work was supported in part by the Grants No.~NSFC-11575089, No.~NSFC-11025525, by The National Key Research and Development Program of China under Grant No. 2017YFA0402200, and by the CAS Center for Excellence in Particle Physics (CCEPP). We are grateful to members of the WeChat group ``QFT fans" for interesting discussions and in particular to Xu Feng for information on lattice calculations. X.-D. Ma would like to thank Li-Lin Yang and Ya Zhang for valuable discussions on the calculation of operator renormalization. We would like to thank the anonymous referee for pointing out an error in eq.~\eqref{eq13} in the previous version and for suggestions that help improve the presentation of our work.

\newpage

\begin{appendices}
\counterwithin{figure}{section}
\section{One-loop contribution with an insertion of effective interactions}
\label{app}

\numberwithin{equation}{section}

\setcounter{equation}{0}

We collect our results for the ultraviolet divergent terms in one-loop diagrams with one insertion of effective interactions due to dim-7 operators that is dressed by SM interactions. Our calculations are done in dimensional regularization ($D=4-2\epsilon$) with minimal subtraction and in $R_\xi$ gauge. The results can be used to extract the anomalous dimension matrix $\gamma$ once a flavor-specified basis is chosen as described in the main text.

We adopt the following shortcuts:
\begin{eqnarray}
\nonumber
\delta &=& 16\pi^2 \epsilon,
\\
\nonumber
W_H&=& {\Tr}\Big[3(Y^\dagger_uY_u)+3(Y^\dagger_dY_d)+(Y^\dagger_eY_e)\Big],
\\
\nonumber
W^1_{pr}&=&\frac{1}{8}\Big[\big(4g_1^2-3g_2^2\big)C^{pr}_{LHD1}
-\big(4g_1^2-15g_2^2\big)C^{rp}_{LHD1}
+4(Y_eY^\dagger_e)_{vp}C^{rv}_{LHD1}
+2(Y_eY^\dagger_e)_{vr}C^{pv}_{LHD1}\Big],
\\
W^2_{pr}&=&\frac{1}{4}\Big[(g_1^2+3g_2^2)C^{pr}_{LHD2}- g_1^2C^{rp}_{LHD2}+\big((Y_eY^\dagger_e)_{vr}C^{pv}_{LHD2}+p\leftrightarrow r\big)\Big],
\end{eqnarray}
where $W_H$ originates from the Higgs field wavefunction renormalization due to Yukawa couplings and $W^{1}_{pr}$ ($W^{2}_{pr}$) appears in insertion of the operator $\mathcal{O}_{LHD1}$ ($\mathcal{O}_{LHD2}$). In the following formulas, $\langle C\calO\rangle$ on the left-hand side stands for an insertion of the effective interaction $C\calO$ with the dim-7 operator $\calO$ and its Wilson coefficient $C$, which yields the one-loop result on the right-hand side due to SM interactions. A subscript $_X$ on the right implies the same field labels as in the left $\langle(C\calO)_X\rangle$.

The results for the operators with $L=2,~B=0$ are:
\begin{eqnarray}
\label{o11l}
\langle(C\calO)_{LH}\rangle\delta&=&
\frac{1}{4}\Big\{\big(3g_1^2+15g_2^2-80\lambda-8W_H\big)C^{pr}_X
+3\Big[(Y_eY^\dagger_e)_{vp}C^{vr}_X+p\leftrightarrow r\Big] \Big\}\mathcal{O}^{pr}_X,~X=LH,
\\
\nonumber
\langle(C\calO)_{LeHD}\rangle\delta&=&
\frac{1}{4}\Big\{\Big[\big(3g_2^2-4\lambda\big)(Y^\dagger_e)_{vr}
-2(Y^\dagger_eY_eY^\dagger_e)_{vr}\Big]C^{pv}_X+p\leftrightarrow r\Big\}\calO^{pr}_{LH}
\\
\nonumber
&&+\frac{1}{4}\Big[\big(3g_1^2-12\lambda-6W_H\big)C^{pr}_X
-(Y_eY^\dagger_e)_{vp}C^{vr}_X-2(Y_e)_{vr}(Y^\dagger_e)_{wp}C^{vw}_X
-8(Y^\dagger_eY_e)_{vr}C^{pv}_X\Big]\mathcal{O}^{pr}_X
\\
&&
-\frac{1}{16}\Big[(Y^\dagger_e)_{vr}C^{pv}_X-p\leftrightarrow r\Big]
\mathcal{O}^{pr}_{LHB}-\frac{1}{4} (Y^\dagger_e)_{vr}C^{pv}_X \mathcal{O}^{pr}_{LHW}
+3(Y^\dagger_d Y_u)_{ps}C^{rt}_X\mathcal{O}^{prst}_{\bar{d}LueH},
~X=LeHD,
\\
\nonumber
\langle(C\calO)_{LHD1}\rangle\delta&=&
\frac{1}{8}\Big\{\Big[3g_2^4C^{pr}_X+4\big(2\lambda(Y_eY^\dagger_e)_{vr}
+(Y_eY^\dagger_eY_eY^\dagger_e)_{vr}\big)C^{pv}_X
-\lambda W^1_{pr}\Big]+p\leftrightarrow r\Big\}\mathcal{O}^{pr}_{LH}
\\
\nonumber
&&
-\frac{1}{4}\Big\{(Y_e)_{vr}\Big[\big(3g_1^2
-4g_2^2\big)C^{pv}_X +\big(3g_1^2+2g_2^2\big)C^{vp}_X \Big]
\\
\nonumber
&&
-\Big[(Y_eY^\dagger_e)_{vp}(Y_e)_{wr}\big(2C^{vw}_X-C^{wv}_X\big)
+4(Y_eY^\dagger_eY_e)_{vr}C^{pv}_X\Big]\Big\}\calO^{pr}_{LeHD}
-(Y_u)_{pr}W^1_{ts}\mathcal{O}^{prst}_{\bar{Q}uLLH}
\\
\nonumber
&&
+\frac{1}{4}\Big\{\Big[2\big(g_1^2-2g_2^2-2W_H\big)C^{pr}_X
+\big(g_1^2-7 g_2^2\big)C^{rp}_X\Big]
\\
\nonumber
&&
-\Big[(Y_eY^\dagger_e)_{vp}\big(7C^{vr}_X-C^{rv}_X\big)
+6(Y_eY^\dagger_e)_{vr}C^{pv}_X\Big]\Big\}\mathcal{O}^{pr}_X
-(Y^\dagger_dY_u)_{ps}(Y_e)_{vt}C^{rv}_X\calO^{prst}_{\bar{d}LueH}
\\
\nonumber
&&
+\frac{1}{2}\Big[g_2^2\big(7C^{pr}_X+C^{rp}_X\big)
+\Big(5(Y_eY^\dagger_e)_{vp}C^{vr}_X+2(Y_eY^\dagger_e)_{vr}C^{pv}_X\Big)\Big]
\mathcal{O}^{pr}_{LHD2}
-(Y^\dagger_dY_u)_{pr}C^{st}_X\mathcal{O}^{prst}_{\bar{d}uLLD}
\\
\nonumber
&&
-\frac{1}{64}\Big\{ \Big[9g_2^2C^{pr}_X+2(Y_eY^\dagger_e)_{vp}\big(2C^{vr}_X-C^{rv}_X\big)\Big]
-p\leftrightarrow r\Big\}\mathcal{O}^{pr}_{LHB}
\\
\nonumber
&&
-\frac{1}{16}\Big[g_2^2\big(7C^{pr}_X
-2C^{rp}_X\big)+2\Big(2(Y_eY^\dagger_e)_{vp}C^{vr}_X
+(Y_eY^\dagger_e)_{vr}C^{pv}_X\Big)\Big]\mathcal{O}^{pr}_{LHW}
\\
\nonumber
&&
-\frac{1}{4}\Big\{3g_1^2(Y^\dagger_e)_{pt}\big(C^{rs}_X+C^{sr}_X\big)
-g_2^2\Big[(Y^\dagger_e)_{pt}\big(C^{rs}_X-C^{sr}_X \big)-2(Y^\dagger_e)_{ps}\big(C^{rt}_X-C^{tr}_X \big)\Big]
\\
\nonumber
&&
-4(Y^\dagger_e)_{pr}W^1_{ts}\Big\}\mathcal{O}^{prst}_{\bar{e}LLLH}
-\frac{1}{2}(Y^\dagger_d)_{ps}
\Big[g_2^2\big(C^{rt}_X-C^{tr}_X\big)
-2\big(W^1_{rt}+W^1_{tr}\big)\Big]\mathcal{O}^{prst}_{\bar{d}LQLH1}
\\
&&
-\frac{1}{12}(Y^\dagger_d)_{ps} \Big[\big(g_1^2-3g_2^2\big)C^{rt}_X+\big(g_1^2+3g_2^2\big)C^{tr}_X
+12W^1_{rt}\Big]\mathcal{O}^{prst}_{\bar{d}LQLH2},~X=LHD1,
\end{eqnarray}
\begin{eqnarray}
\nonumber
\langle(C\calO)_{LHD2}\rangle\delta&=&
\frac{1}{16}\Big\{\Big[3\big(g_1^4+2g_1^2g_2^2+3g_2^4\big)C^{pr}_X
+8\big(2\lambda(Y_eY^\dagger_e)_{vr}+(Y_eY^\dagger_eY_eY^\dagger_e)_{vr}\big)
C^{pv}_X-\lambda W^2_{pr}\Big]
+p\leftrightarrow r\Big\}\mathcal{O}^{pr}_{LH}
\\
\nonumber
&&
-\frac{1}{16}\Big[\big( 13g_1^2-17g_2^2-8\lambda\big)(Y_e)_{vr}C^{pv}_X
-4\Big(5(Y_eY^\dagger_eY_e)_{vr}C^{pv}_X
+(Y_eY^\dagger_e)_{vp}(Y_e)_{wr}C^{vw}_X\Big)
\Big]\mathcal{O}^{pr}_{LeHD}
\\
\nonumber
&&
+\frac{1}{8}\Big\{\Big[\big(7g^2_1+11 g^2_2+8\lambda\big)C^{pr}_X
+4\big(g^2_1+2 g^2_2\big)C^{rp}_X\Big]
\\
\nonumber
&&
+4\Big[(Y_eY^\dagger_e)_{vp}\big(C^{vr}_X-C^{rv}_X\big)
-(Y_eY^\dagger_e)_{vr}C^{pv}_X\Big]\Big\}\mathcal{O}^{pr}_{LHD1}
\\
\nonumber
&&
-\frac{1}{4}\Big\{\big(5g_1^2-g_2^2+8\lambda+4W_H\big)C^{pr}_X
+\big(3g_1^2+7g_2^2\big)C^{rp}_X-3\Big[(Y_eY^\dagger_e)_{vp}C^{vr}_X
+(Y_eY^\dagger_e)_{vr}C^{pv}_X\Big]\Big\}\mathcal{O}^{pr}_X
\\
\nonumber
&&
-\frac{1}{12}(Y^\dagger_d)_{ps} \Big( \big(g_1^2-9g_2^2\big)C^{rt}_X
+12W^2_{rt}\Big)\mathcal{O}^{prst}_{\bar{d}LQLH2}
+\frac{3}{128}\big(g_1^2-g_2^2\big)\big(C^{pr}_X-C^{rp}_X\big)\calO^{pr}_{LHB}
\\
\nonumber
&&
+\frac{1}{32}\Big\{\Big[\big(3g_1^2-7g_2^2\big)C^{pr}_X
-4g_2^2C^{rp}_X\Big]-4\Big[(Y_eY^\dagger_e)_{vr}C^{pv}_X
+p \leftrightarrow r\Big]\Big\}\mathcal{O}^{pr}_{LHW}
\\
\nonumber
&&
-\frac{1}{4}\Big\{3\big(g_1^2-g_2^2\big)\Big[(Y^\dagger_e)_{pt}C^{rs}_X
-(Y^\dagger_e)_{ps}\big(C^{rt}_X-C^{tr}_X\big)\Big]
-4(Y^\dagger_e)_{pr}W^2_{ts}\Big\}\mathcal{O}^{prst}_{\bar{e}LLLH}
\\
\nonumber
&&
+\frac{1}{12}(Y^\dagger_d)_{ps}
\Big[\big(g_1^2-9g_2^2\big)\big(C^{rt}_X-C^{tr}_X\big)
-12\big(W^2_{rt}+W^2_{tr}\big)\Big]\mathcal{O}^{prst}_{\bar{d}LQLH1}
\\
&&-2(Y^\dagger_dY_u)_{ps}(Y_e)_{vt}C^{rv}_X\mathcal{O}^{prst}_{\bar{d}LueH}
- (Y_u)_{pr}W^2_{ts}\mathcal{O}^{prst}_{\bar{Q}uLLH}
-\frac{1}{2}(Y^\dagger_dY_u)_{pr}C^{st}_X\mathcal{O}^{prst}_{\bar{d}uLLD},
~X=LHD2,
\\
\nonumber
\langle(C\calO)_{LHB}\rangle\delta&=&
\frac{1}{4}\Big\{\big(g_1^2+10g_2^2-8\lambda-4W_H\big)C^{pr}_X
+3\Big[(Y_eY^\dagger_e)_{vp}C^{vr}_X-p\leftrightarrow r\Big]\Big\}
\calO^{pr}_X-\frac{3}{2}g_1^2C^{pr}_X\mathcal{O}^{pr}_{LHW}
\\
\nonumber
&&
-3g^2_1\Big[(Y_e^\dagger)_{pr}C^{st}_X
+(Y_e^\dagger)_{pt}C^{rs}_X-4(Y_e^\dagger)_{ps}C^{rt}_X\Big]
\mathcal{O}^{prst}_{\bar{e}LLLH}
+\frac{4}{3}g^2_1(Y_d^\dagger)_{ps}C^{rt}_X\calO^{prst}_{\bar{d}LQLH1}
\\
&&
-\frac{2}{3}g^2_1(Y_d^\dagger)_{ps}C^{rt}_X\calO^{prst}_{\bar{d}LQLH2},
~X=LHB,
\\
\nonumber
\langle(C\calO)_{LHW}\rangle\delta&=&
\frac{3}{4}g^2_2 \Big\{\Big[g^2_2C^{pr}_X+2(Y_eY^\dagger_e)_{vr}C^{pv}_X\Big]
+p\leftrightarrow r\Big\}\mathcal{O}^{pr}_{LH}
-\frac{3}{8}g_2^2\big(C^{pr}_X-C^{rp}_X\big)\calO^{pr}_{LHB}
\\
\nonumber
&&+\frac{1}{8}\Big\{\big(8g_1^2-9g_2^2
-16\lambda-8W_H\big)C^{pr}_X-17g_2^2C^{rp}_X
\\
\nonumber
&&+2\Big[(Y_eY^\dagger_e)_{vp}\big(3C^{vr}_X-4C^{rv}_X\big)
-9(Y_eY^\dagger_e)_{vr}C^{pv}_X\Big]\Big\}\mathcal{O}^{pr}_X
\\
\nonumber
&&+\frac{1}{2}g^2_2
\Big\{(Y^\dagger_e)_{pt}\big(5C^{rs}_X+C^{sr}_X\big)
-4(Y^\dagger_e)_{ps}\big(C^{rt}_X-C^{tr}_X\big)\Big\}
\calO^{prst}_{\bar{e}LLLH}
\\
&&
-2g^2_2(Y_d^\dagger)_{ps}\big(C^{rt}_X-C^{tr}_X\big)
\calO^{prst}_{\bar{d}LQLH1}
+\frac{1}{2}g^2_2 (Y_d^\dagger)_{ps}\big(5C^{rt}_X+C^{tr}_X\big) \mathcal{O}^{prst}_{\bar{d}LQLH2},~X=LHW,
\\
\nonumber
\langle(C\calO)_{\bar{e}LLLH}\rangle\delta
&=&\frac{1}{2}\Big[\lambda(Y_e)_{vw}-(Y_eY^\dagger_e Y_e)_{vw}\Big]
\Big[\big(2C^{wvpr}_X+C^{wpvr}_X\big)+p\leftrightarrow r\Big]\mathcal{O}^{pr}_{LH}
+\frac{3}{16}(Y_e)_{vw}\big(C^{wpvr}_X
-C^{wrvp}_X\big)\mathcal{O}^{pr}_{LHB}
\\
\nonumber
&&
+\frac{1}{8}\Big\{\big(9g_1^2+7g_2^2-4W_H\big)C^{prst}_X
-8g_2^2\big(C^{prts}_X-C^{pstr}_X+2C^{ptsr}_X\big)
\\
\nonumber
&&
-4(Y^\dagger_e)_{pr}(Y_d)_{vw}\big(2C^{wvst}_X+C^{wsvt}_X
+C^{wstv}_X-C^{wtsv}_X\big)
-2\Big[6(Y^\dagger_e Y_e)_{pv}C^{vrst}_X
\\
\nonumber
&&
+(Y_e Y_e^\dagger)_{vr}\big(C^{pvst}_X+4C^{pvts}_X\big)+5(Y_e Y_e^\dagger)_{vs}C^{prvt}_X
-(Y_e Y_e^\dagger)_{vt}\big(4C^{prvs}_X+3C^{prsv}_X\big)
\Big]\Big\}\mathcal{O}^{prst}_X
\\
\nonumber
&&
+\frac{1}{2}(Y^\dagger_d)_{ps}(Y_e)_{vw}\Big\{
\Big(2C^{wvtr}_X+C^{wtvr}_X+C^{wtrv}_X-C^{wrtv}_X\Big)
\calO^{prst}_{\bar{d}LQLH2}
\\
\nonumber
&&
-\Big[\big(2C^{wvtr}_X
+C^{wtvr}_X\big)+r\leftrightarrow t\Big]\mathcal{O}^{prst}_{\bar{d}LQLH1}\Big\}
+\frac{1}{4}(Y_e)_{vw}\big(C^{wrvp}_X+C^{wprv}_X+C^{wrpv}_X\big)
\mathcal{O}^{pr}_{LHW}
\\
&&
+\frac{1}{2}(Y_u)_{pr}(Y_e)_{vw}\Big(2C^{wvst}_X+C^{wsvt}_X
+C^{wstv}_X-C^{wtsv}_X\Big)\mathcal{O}^{prst}_{\bar{Q}uLLH},
~X=\bar{e}LLLH,
\\
\nonumber
\langle(C\calO)_{\bar{d}LueH}\rangle\delta&=&
3(Y^\dagger_uY_d)_{vw}C^{wpvr}_X\mathcal{O}^{pr}_{LeHD}
+(Y^\dagger_u)_{vs}(Y^\dagger_e)_{wt}C^{prvw}_X\calO^{prst}_{\bar{d}LQLH1}
-(Y^\dagger_u)_{vs}(Y^\dagger_e)_{wt}C^{prvw}_X\calO^{prst}_{\bar{d}LQLH2}
\\
\nonumber
&&
-\frac{1}{4}\Big[6(Y^\dagger_dY_d)_{pv}C^{vrst}_X
-3(Y_eY^\dagger_e)_{vr}C^{pvst}_X+6(Y^\dagger_uY_u)_{vs}C^{prvt}_X
+4(Y^\dagger_eY_e)_{vt}C^{prsv}_X
+2(Y_e)_{vt}(Y^\dagger_e)_{wr}
C^{pvsw}_X\Big]
\\
&&
\times\mathcal{O}^{prst}_X
+\frac{1}{8}\big(23g_1^2+9g_2^2-4W_H\big)C^{prst}_X\mathcal{O}^{prst}_X
-\frac{1}{2}(Y_d)_{pv}(Y^\dagger_e)_{ws}C^{vtrw}_X\calO^{prst}_{\bar{Q}uLLH},
~X=\bar{d}LueH,
\label{examplesum}
\end{eqnarray}

\begin{eqnarray}
\nonumber
\langle(C\calO)_{\bar{d}LQLH1}\rangle\delta&=&
\frac{3}{2}\big[\lambda(Y_d)_{vw}-(Y_dY^\dagger_d Y_d)_{vw}\big]\big(C^{wpvr}_X+C^{wrvp}_X\big)\mathcal{O}^{pr}_{LH}
+\frac{1}{16}(Y_d)_{vw}\big(C^{wpvr}_X-C^{wrvp}_X\big)\mathcal{O}^{pr}_{LHB}
\\
\nonumber
&&
+\frac{3}{4}(Y_d)_{vw}C^{wrvp}_X\calO^{pr}_{LHW}
-\frac{3}{2}(Y^\dagger_e)_{pr}(Y_d)_{vw}C^{wsvt}_X\calO^{prst}_{\bar{e}LLLH}
\\
\nonumber
&&
+\frac{1}{72}\Big\{\big(41g_1^2+63g_2^2+96g_3^2-36W_H\big)C^{prst}_X
-16\big(5g_1^2+9g_2^2-12g_3^2\big)C^{ptsr}_X
\\
\nonumber
&&
-18\Big[6(Y^\dagger_dY_d)_{pv}C^{vrst}_X
+(Y_uY_u^\dagger+5Y_dY_d^\dagger )_{vs}C^{prvt}_X
+(Y_eY_e^\dagger )_{vr}
C^{pvst}_X -3(Y_eY^\dagger_e)_{vt}C^{prsv}_X
\\
\nonumber
&&
+\frac{3}{2}(Y^\dagger_d)_{ps}(Y_d)_{vw}
\big(C^{wtvr}_X+C^{wrvt}_X\big)\Big]\Big\}\calO^{prst}_X
-(Y_u)_{vs}(Y_e)_{wt}\big(C^{pwvr}_X-C^{prvw}_X\big)\calO^{prst}_{\bar{d}LueH}
\\
\nonumber
&&
+\frac{1}{9}\Big[\big(10g_1^2+9g_2^2
-24g_3^2\big)C^{ptsr}_X-9g_2^2C^{prst}_X\Big]
\calO^{prst}_{\bar{d}LQLH2}
\\
\nonumber
&&
-\frac{1}{2}\Big[2(Y_eY^\dagger_e)_{vr}C^{pvst}_X
-2(Y_dY^\dagger_d)_{vs}C^{prvt}_X+(Y_uY^\dagger_u)_{vs}
C^{prvt}_X-3(Y^\dagger_d)_{ps}(Y_d)_{vw}C^{wtvr}_X \Big]\mathcal{O}^{prst}_{\bar{d}LQLH2}
\\
&&
+\frac{1}{2}\Big[3(Y_u)_{pr}(Y_d)_{vw}-(Y_d)_{pw}(Y_u)_{vr}\Big]
C^{wsvt}_X\mathcal{O}^{prst}_{\bar{Q}uLLH},
~X=\bar{d}LQLH1,
\\
\nonumber
\langle(C\calO)_{\bar{d}LQLH2}\rangle\delta&=&
\frac{3}{4}(Y_d)_{vw}\big(C^{wpvr}_X+C^{wrvp}_X\big)\calO^{pr}_{LHW}
-\frac{3}{2}(Y^\dagger_e)_{pr}(Y_d)_{vw}\big(C^{wsvt}_X-C^{wtvs}_X\big)
\calO^{prst}_{\bar{e}LLLH}
\\
\nonumber
&&
-\Big\{g_2^2\big(2C^{prst}_X+C^{ptsr}_X\big)
+\Big[(Y_eY^\dagger_e)_{vr}C^{pvst}_X-(Y_eY^\dagger_e)_{vt}C^{prsv}_X
\Big]\Big\}\mathcal{O}^{prst}_{\bar{d}LQLH1}
\\
\nonumber
&&
+\frac{1}{72}\Big\{\big(41g_1^2+207g_2^2+96g_3^2
-36W_H\big)C^{prst}_X
+8\big(10g_1^2-9g_2^2-24g_3^2\big)C^{ptsr}_X
\\
\nonumber
&&
-18\Big[6(Y^\dagger_dY_d)_{pv}C^{vrst}_X
+\big(5(Y_uY^\dagger_u)_{vs}
-3(Y_dY^\dagger_d)_{vs}\big)C^{prvt}_X
+(Y_eY^\dagger_e)_{vr}C^{pvst}_X
+5(Y_eY^\dagger_e)_{vt}C^{prsv}_X
\\
\nonumber
&&
-6(Y^\dagger_d)_{ps}(Y_d)_{vw}\big(C^{wtvr}_X
-C^{wrvt}_X\big)\Big]\Big\}\mathcal{O}^{prst}_X
-\frac{1}{2}\Big[(Y_e)_{vt}(Y_u)_{ws}C^{pvwr}_X
+2(Y_u)_{vs}(Y_e)_{wt}C^{prvw}_X\Big]\mathcal{O}^{prst}_{\bar{d}LueH}
\\
&&+\frac{1}{2}\Big[3(Y_u)_{pr}(Y_d)_{vw}-(Y_d)_{pw}(Y_u)_{vr}\Big]
\big(C^{wsvt}_X-C^{wtvs}_X\big)\calO^{prst}_{\bar{Q}uLLH},
~X=\bar{d}LQLH2,
\\
\nonumber
\langle(C\calO)_{\bar{Q}uLLH}\rangle\delta&=&
-3\Big[\lambda(Y^\dagger_u)_{vw}-(Y^\dagger_u Y_uY^\dagger_u)_{vw}\Big]\big(C^{wvpr}_X+C^{wvrp}_X\big)\calO^{pr}_{LH}
+3(Y^\dagger_e)_{pr}(Y^\dagger_u)_{vw}C^{wvst}_X\calO^{prst}_{\bar{e}LLLH}
\\
\nonumber
&&
+\Big[3(Y^\dagger_d)_{ps}
(Y^\dagger_u)_{vw}-(Y^\dagger_d)_{pw}(Y^\dagger_u)_{vs}\Big]
\big(C^{wvtr}_X+C^{wvrt}_X\big)\mathcal{O}^{prst}_{\bar{d}LQLH1}
\\
\nonumber
&&
-\Big[3(Y^\dagger_d)_{ps}(Y^\dagger_u)_{vw} -(Y^\dagger_d)_{pw}(Y^\dagger_u)_{vs}\Big]C^{wvtr}_X
\mathcal{O}^{prst}_{\bar{d}LQLH2}
-\frac{1}{2}(Y^\dagger_d)_{pv}(Y_e)_{wt}
\big(2C^{vswr}_X+C^{vsrw}_X\big)\mathcal{O}^{prst}_{\bar{d}LueH}
\\
\nonumber
&&
-\frac{1}{24}\bigg\{\big(g_1^2-45g_2^2-96g_3^2+12W_H\big)C^{prst}_X
+36g_2^2C^{prts}_X
\\
\nonumber
&&
+6\Big[5(Y_uY^\dagger_u)_{pv}C^{vrst}_X
+6(Y^\dagger_uY_u)_{vr}C^{pvst}_X
+5(Y_eY^\dagger_e)_{vs}C^{prvt}_X
-(Y_eY^\dagger_e)_{vt}\big(4C^{prvs}_X
+3C^{prsv}_X\big)\Big]
\\
&&
+6\Big((Y_dY^\dagger_d)_{pv}\big(3C^{vrst}_X-2C^{vrts}_X\big)
+12(Y_u)_{pr}(Y^\dagger_u)_{vw}C^{wvst}_X\Big)\bigg\} \mathcal{O}^{prst}_X,
~X=\bar{Q}uLLH,
\\
\nonumber
\label{o121l}
\langle(C\calO)_{\bar{d}uLLD}\rangle\delta&=&
-6(Y^\dagger_uY_d)_{vw}C^{wvpr}_X\mathcal{O}^{pr}_{LHD1}
+(Y^\dagger_u)_{vs} \Big\{g^2_2\big(C^{pvrt}_X+2C^{pvtr}_X\big)
+\Big[(Y_eY^\dagger_e)_{wr}C^{pvtw}_X+r\leftrightarrow t\Big]\Big\}\mathcal{O}^{prst}_{\bar{d}LQLH1}
\\
\nonumber
&&
-\frac{1}{6}(Y^\dagger_u)_{vs}\Big[g^2_1\big(2C^{pvrt}_X-C^{pvtr}_X\big)
+3g^2_2\big(C^{pvrt}_X+2C^{pvtr}_X\big)
+6(Y_eY^\dagger_e)_{wt}C^{pvrw}_X\Big]\mathcal{O}^{prst}_{\bar{d}LQLH2}
\\
\nonumber
&&
+\frac{1}{12}\Big\{(Y_e)_{vt}\big(19g_1^2-3g_2^2\big)
\big(C^{psrv}_X+C^{psvr}_X\big)
-6(Y_e)_{wt}\Big[(Y^\dagger_dY_d)_{pv}C^{vsrw}_X
+(Y^\dagger_uY_u)_{vs}C^{pvrw}_X\Big]
\\
\nonumber
&&
+6(Y_e)_{vt}(Y_eY^\dagger_e)_{wr}\big(C^{psvw}_X+C^{pswv}_X\big)  \Big\}\mathcal{O}^{prst}_{\bar{d}LueH}
-\frac{1}{4}(Y_d)_{pv}\Big[g^2_1\big(C^{vrst}_X-C^{vrts}_X\big)
+3g^2_2\big(C^{vrst}_X+C^{vrts}_X\big)
\\
\nonumber
&&
+4(Y_eY^\dagger_e)_{ws}C^{vrtw}_X\Big]\mathcal{O}^{prst}_{\bar{Q}uLLH}
-\frac{1}{12}\Big[9\big(g_2^2+g_1^2\big)C^{prst}_X-2 \big(3g_2^2+2g_1^2\big)\big(C^{prst}_X+C^{prts}_X\big)\Big]\calO^{prst}_X
\\
&&
-\frac{1}{4}\Big[2(Y_d^\dagger Y_d)_{pv}C^{vrst}_X+2(Y_u^\dagger Y_u)_{vr}C^{pvst}_X+(Y_eY_e^\dagger )_{vs}C^{prvt}_X  +(Y_eY_e^\dagger )_{vt}C^{prsv}_X\Big]\mathcal{O}^{prst}_X,
~X=\bar{d}uLLD.
\end{eqnarray}
For completeness, we reproduce the results for the operators with $B=-L=1$ that were obtained in Ref.~\cite{Liao:2016hru}:
\begin{eqnarray}
\nonumber
\langle(C\calO)_{\bar{L}dud\tilde{H}}\rangle\delta&=&
	\Big\{\Big[2g^2_3+\frac{9}{8}g^2_2 +\frac{17}{24}g^2_1-\frac{1}{2}W_H\Big]C^{prst}_X
+\frac{5}{3}g^2_1C^{ptsr}_X
+\frac{3}{4}(Y_eY^\dagger_e)_{pv}C^{vrst}_X
-\frac{3}{2}(Y^\dagger_dY_d)_{vr}C^{pvst}_X
	\\
\nonumber
&&-\frac{3}{2}(Y^\dagger_dY_d)_{vt}C^{prsv}_X
-(Y^\dagger_uY_u)_{vs}C^{prvt}_X\Big\} \calO^{prst}_X
+\frac{1}{4}(Y^\dagger_e)_{pv}(Y^\dagger_u)_{wr}
\big(C^{vtws}_X-C^{vswt}_X\big)
\calO^{prst}_{\bar{e}Qdd\tilde{H}}
\\
\nonumber
&&
+\frac{1}{8}\Big\{\Big[(Y^\dagger_uY_d)_{vs}
\big(C^{prvt}_X+C^{ptvr}_X\big)+(Y^\dagger_uY_d)_{vr}C^{psvt}_X\Big]
-s\leftrightarrow t\Big\}\calO^{prst}_{\bar{L}dddH}
\\
&&+\Big\{\Big[(Y^\dagger_u)_{vs}(Y^\dagger_d)_{wt}C^{prvw}_X
+\frac{1}{2}(Y^\dagger_d)_{vs}(Y^\dagger_u)_{wt}C^{pvwr}_X\Big]
+s\leftrightarrow t\Big\}\calO^{prst}_{\bar{L}dQQ\tilde{H}},
~X=\bar{L}dud\tilde{H},
\\
\nonumber
\langle(C\calO)_{\bar{L}dddH}\rangle\delta&=&
(Y^\dagger_dY_u)_{vs}\big(C^{pvrt}_X
+C^{prvt}_X\big)\calO^{prst}_{\bar{L}dud\tilde{H}}
	+\Big\{\Big[2g^2_3+ \frac{9}{8}g^2_2+\frac{13}{24}g^2_1-\frac{1}{2}W_H\Big]C^{prst}_X
	\\
	&&
-\Big[\frac{5}{4}(Y_eY^\dagger_e)_{pv}C^{vrst}_X
+(Y^\dagger_dY_d)_{vr}
C^{pvst}_X+(Y^\dagger_dY_d)_{vs}C^{prvt}_X
+(Y^\dagger_dY_d)_{vt}C^{prsv}_X\Big]\Big\}\calO^{prst}_X,
~X=\bar{L}dddH,
\\
\nonumber
\langle(C\calO)_{\bar{e}Qdd\tilde{H}}\rangle\delta &=&
\Big\{\Big[2g^2_3+\frac{9}{8}g^2_2 -\frac{11}{24}g^2_1-\frac{1}{2}W_H\Big]C^{prst}_X
-\Big[(Y^\dagger_eY_e)_{pv}C^{vrst}_X
+\frac{1}{4}(-3Y_uY^\dagger_u+5Y_dY^\dagger_d)_{vr} C^{pvst}_X
	\\
\nonumber
	&&+\frac{3}{2}(Y^\dagger_dY_d)_{vs}C^{prvt}_X
+\frac{3}{2}(Y^\dagger_dY_d)_{vt}C^{prsv}_X\Big]
+\frac{1}{2}(Y^\dagger_d)_{wr}\Big((Y_d)_{vs}C^{pvwt}_X
+(Y_d)_{vt}C^{pvsw}_X\Big)\Big\}\calO^{prst}_X
	\\
	&&
-2(Y_e)_{pv}(Y_u)_{ws}C^{vwrt}_X\calO^{prst}_{\bar{L}dud\tilde{H}}
+(Y_e)_{pv}(Y^\dagger_d)_{ws}C^{vtwr}_X
\calO^{prst}_{\bar{L}dQQ\tilde{H}},~X=\bar{e}Qdd\tilde{H},
\\
\nonumber
\langle(C\calO)_{\bar{L}dQQ\tilde{H}}\rangle\delta&=&
(Y_u)_{vs}(Y_d)_{wt}\big(C^{prvw}_X+C^{prwv}_X\big)
\calO^{prst}_{\bar{L}dud\tilde{H}}
+\frac{1}{4}(Y^\dagger_e)_{pv}\Big[(Y_d)_{ws}\big(2C^{vtwr}_X-C^{vtrw}_X\big)
-s\leftrightarrow t\Big]\calO^{prst}_{\bar{e}Qdd\tilde{H}}
	\\
\nonumber
&&
+\Big\{\Big[2g^2_3+\frac{15}{8}g^2_2+\frac{19}{24}g^2_1
-\frac{1}{2}W_H\Big]C^{prst}_X
+ \frac{3}{2}g^2_2C^{prts}_X
+\frac{1}{2}(Y_eY^\dagger_e)_{pv}C^{vrts}_X
	\\
\nonumber
	&&
+\frac{3}{4}(Y_uY^\dagger_u)_{vt}C^{prsv}_X
-\frac{5}{4}(Y_uY^\dagger_u)_{vs}C^{prvt}_X
-(Y_uY^\dagger_u)_{vt}C^{prvs}_X
\\
\nonumber
&&-\frac{3}{2}(Y^\dagger_dY_d)_{vr}C^{pvst}_X
+\frac{1}{4}(Y_dY^\dagger_d)_{vt}\big(2C^{prvs}_X-5C^{prsv}_X\big)
-\frac{1}{4}(Y_dY^\dagger_d)_{vs}C^{prvt}_X
	\\
&&
+\frac{1}{2}(Y_d)_{wr}\Big((Y^\dagger_d)_{vs}C^{pvwt}_X
+(Y^\dagger_d)_{vt}C^{pvsw}_X\Big)\Big\}\calO^{prst}_X,~
X=\bar{L}dQQ\tilde{H},
\\
\nonumber
\langle(C\calO)_{\bar{L}QddD}\rangle\delta&=&
	\Big\{\Big[\frac{1}{18}g^2_1\big(11C^{pvrt}_X
-13C^{pvtr}_X\big) +\frac{4}{3}g^2_3\big(C^{pvrt}_X
-2C^{pvtr}_X\big)\Big](Y_u)_{vs}
+(Y_d)_{vr}(Y_d^\dagger Y_u)_{ws}C^{pvtw}_X
	\\
\nonumber
	&&			+(Y_u)_{vs}\Big[( Y_d^\dagger Y_d)_{wt}C^{pvrw}_X-r\leftrightarrow t\Big]
	-\frac{1}{2}(Y^\dagger_dY_u)_{ws}\Big[2(Y_d)_{vr}C^{pvtw}_X
	-(Y_d)_{vt}C^{pvrw}_X\Big]
	\Big\}\calO^{prst}_{\bar{L}dud\tilde{H}}
	\\
\nonumber
	&&
	+\Big\{\Big[\Big( \frac{1}{12}g^2_1(Y_d)_{vs}C^{pvrt}_X
-\frac{1}{9}(g^2_1-6g_3^2)(Y_d)_{vr}C^{pvst}_X
+\frac{1}{4}(Y_d)_{vt}(Y^\dagger_dY_d)_{wr}C^{pvsw}_X\Big) +r\leftrightarrow t\Big]	
	\\
\nonumber
	&&
	-s\leftrightarrow t\Big\}\calO^{prst}_{\bar{L}dddH}
	+\frac{1}{2}(Y^\dagger_e)_{pv}\Big[ \Big(g^2_1C^{vrst}_X
+(Y_d^\dagger Y_d)_{wt}C^{vrsw}_X
	\Big)-s\leftrightarrow t\Big]O^{prst}_{\bar{e}Qdd\tilde{H}}
	\\
\nonumber
	&&
	+\frac{1}{18}\Big\{(g_1^2-24g^2_3)(Y^\dagger_d)_{vt}
	\big(C^{psrv}_X+C^{psvr}_X\big)
	-9\Big[(Y_d^\dagger Y_d)_{vr}(Y_d^\dagger)_{wt}\big(C^{psvw}_X+C^{pswv}_X\big)
	\\
\nonumber
	&&
+\Big((Y_uY_u^\dagger)_{vs}(Y_d^\dagger)_{wt}
+s\leftrightarrow t\Big)C^{pvrw}_X
	-(Y_eY_e^\dagger )_{pv}(Y_d^\dagger)_{ws}C^{vtrw}_X
\Big]\Big\}
\calO^{prst}_{\bar{L}dQQ\tilde{H}}
	\\
\nonumber
	&&
+\frac{1}{18}\Big\{g^2_1\big(C^{prts}_X
-5C^{prst}_X\big)+12g^2_3\big(C^{prst}_X-2C^{prts}_X\big)
+9(Y^\dagger_d)_{wr}\Big[(Y_d)_{vs}C^{pvwt}_X
+(Y_d)_{vt}C^{pvsw}_X\Big]
	\\
\nonumber
&&
-\frac{9}{2}\Big[2\Big((Y_dY^\dagger_d)_{vr}C^{pvst}_X
+(Y^\dagger_dY_d)_{vs}C^{prvt}_X
+(Y^\dagger_dY_d)_{vt}C^{prsv}_X\Big)
\\
&&
+(Y_eY^\dagger_e)_{pv}C^{vrst}_X
+(Y_uY^\dagger_u)_{vr}C^{pvst}_X
\Big]\Big\}\calO^{prst}_X
+(Y^\dagger_e)_{pv}(Y_d)_{wr}C^{vwst}_X
\calO^{prst}_{\bar{e}dddD},~X=\bar{L}QddD,
\end{eqnarray}
\begin{eqnarray}
\nonumber
\langle(C\calO)_{\bar{e}dddD}\rangle\delta&=&
-\frac{1}{2}(Y_e)_{pv}(Y^\dagger_dY_u)_{ws}C^{vrtw}_X
\calO^{prst}_{\bar{L}dud\tilde{H}}
	-\frac{1}{24}\Big\{\Big[4g^2_1(Y_e)_{pv}\big( C^{vrst}_X+C^{vsrt}_X\big)
	+3(Y_e)_{pv}\Big( (Y^\dagger_dY_d)_{wt}C^{vrsw}_X
	\\
\nonumber
	&&
	+(Y^\dagger_dY_d)_{wt}C^{vsrw}_X
	+(Y^\dagger_dY_d)_{wr}C^{vtsw}_X\Big)\Big]-s\leftrightarrow t \Big\}\calO^{prst}_{\bar{L}dddH}
	\\
\nonumber
	&&
-\frac{1}{12}\Big\{\Big[(g^2_1+12g_3^2)(Y^\dagger_d)_{vr}
\big(C^{pvst}_X+C^{psvt}_X+C^{pstv}_X\big)
	-3\Big((Y_d^\dagger Y_d)_{vs}(Y_d^\dagger)_{wr}+(Y_d^\dagger Y_d)_{ws}(Y_d^\dagger )_{vr}\Big) C^{ptvw}_X
	\\
\nonumber
&&
	+3\Big(2(Y_d^\dagger )_{vr}(Y_d^\dagger Y_d)_{wt}-(Y_d^\dagger)_{wr}(Y_d^\dagger Y_d)_{vt}\Big)C^{pvsw}_X\Big]
	-s\leftrightarrow t\Big\}\calO^{prst}_{\bar{e}Qdd\tilde{H}}
	\\
\nonumber
&&
-\frac{1}{2}(Y_e)_{pv}(Y^\dagger_d)_{ws}(Y^\dagger_d)_{xt}
\big(C^{vrwx}_X+C^{vxrw}_X
+C^{vrxw}_X\big)\calO^{prst}_{\bar{L}dQQ\tilde{H}}
	\\
\nonumber
&&
+\frac{1}{2}(Y_e)_{pv}(Y^\dagger_d)_{wr}\big(C^{vwst}_X
+C^{vswt}_X+C^{vstw}_X\big)\calO^{prst}_{\bar{L}QddD}
	\\
\nonumber
	&&
+\frac{1}{18}\Big\{2g^2_1\Big[ \frac{2}{3}C^{prst}_X+C^{prts}_X+C^{psrt}_X+C^{pstr}_X
+C^{ptrs}_X+C^{ptsr}_X\Big]
	\\
\nonumber
&&
	+12g^2_3\Big[2C^{prst}_X-C^{psrt}_X-C^{prts}_X-C^{pstr}_X
	-C^{ptrs}_X-C^{ptsr}_X\Big]
\\
&&
-9\Big[(Y^\dagger_eY_e)_{pv}C^{vrst}_X
+(Y^\dagger_dY_d)_{vr}C^{pvst}_X
+(Y^\dagger_dY_d)_{vs}C^{prvt}_X
+(Y^\dagger_dY_d)_{vt}C^{prsv}_X\Big]\Big\}
\calO^{prst}_X,~X=\bar{e}dddD.
\end{eqnarray}

\begin{figure}[ht!]
\centering
\includegraphics[width=0.9\linewidth]{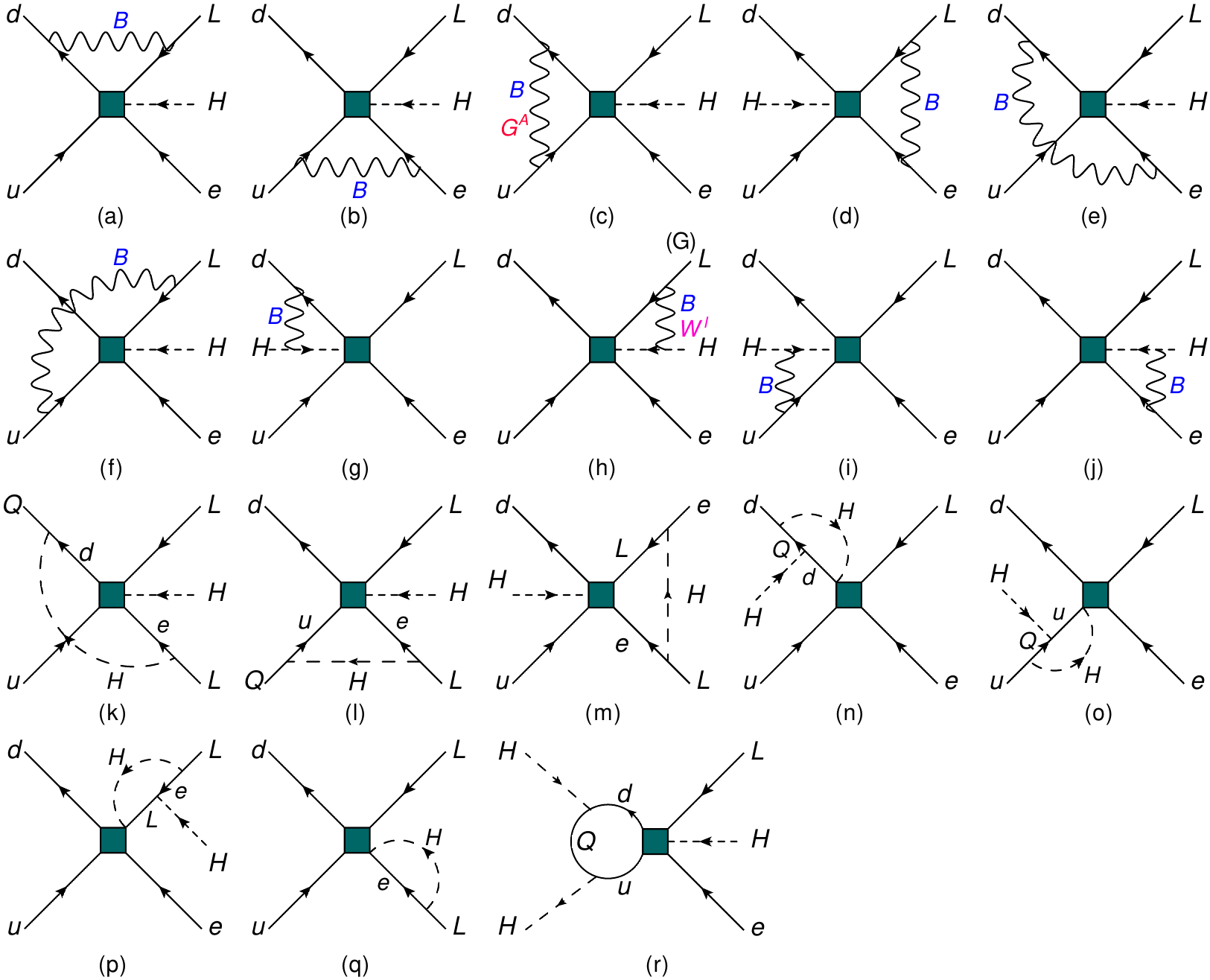}
\caption{One-loop Feynman diagrams with an insertion of the effective interaction $(C\mathcal{O})_{\bar{d}LueH}$.}
\label{fig3}
\end{figure}

As an illustration of our detailed calculation, we show the one-loop correction with an insertion of the effective interaction $(C\mathcal{O})_{\bar{d}LueH}$. All one particle irreducible  divergent Feynman diagrams are shown Fig.~\ref{fig3}. We verified that diagrams (q) and (r) contain a derivative that combines with additional diagrams obtained by attaching a gauge field to an internal propagator of those two diagrams to form a gauge covariant derivative. The result is, diagram by diagram,
\begin{eqnarray}
\nonumber
\langle (C\mathcal{O})_{\bar{d}LueH}\rangle^{(a)}\delta&=&(\xi_1+3)y_dy_Lg_1^2C_{\bar{d}LueH}^{prst}\mathcal{O}_{\bar{d}LueH}^{prst},
\\
\nonumber
\langle (C\mathcal{O})_{\bar{d}LueH}\rangle^{(b)}\delta&=&-(\xi_1+3)y_uy_eg_1^2C_{\bar{d}LueH}^{prst}\mathcal{O}_{\bar{d}LueH}^{prst},
\\
\nonumber
\langle (C\mathcal{O})_{\bar{d}LueH}\rangle^{(c)}\delta&=&\Big(\frac{4}{3}\xi_3g_3^2+\xi_1 y_dy_ug_1^2\Big)C_{\bar{d}LueH}^{prst}\mathcal{O}_{\bar{d}LueH}^{prst},
\\
\nonumber
\langle (C\mathcal{O})_{\bar{d}LueH}\rangle^{(d)}\delta&=&-\xi_1 y_Ly_eg_1^2C_{\bar{d}LueH}^{prst}\mathcal{O}_{\bar{d}LueH}^{prst},
\\
\nonumber
\langle (C\mathcal{O})_{\bar{d}LueH}\rangle^{(e)}\delta&=&\xi_1 y_dy_eg_1^2C_{\bar{d}LueH}^{prst}\mathcal{O}_{\bar{d}LueH}^{prst},
\\
\nonumber
\langle (C\mathcal{O})_{\bar{d}LueH}\rangle^{(f)}\delta&=&-\xi_1 y_Ly_ug_1^2C_{\bar{d}LueH}^{prst}\mathcal{O}_{\bar{d}LueH}^{prst},
\\
\nonumber
\langle (C\mathcal{O})_{\bar{d}LueH}\rangle^{(g)}\delta&=&\xi_1 y_dy_Hg_1^2C_{\bar{d}LueH}^{prst}\mathcal{O}_{\bar{d}LueH}^{prst},
\\
\nonumber
\langle (C\mathcal{O})_{\bar{d}LueH}\rangle^{(h)}\delta&=&\Big(\frac{3}{4}\xi_2g_2^2-\xi_1 y_Ly_Hg_1^2\Big)C_{\bar{d}LueH}^{prst}\mathcal{O}_{\bar{d}LueH}^{prst},
\\
\nonumber
\langle (C\mathcal{O})_{\bar{d}LueH}\rangle^{(i)}\delta&=&-\xi_1 y_uy_Hg_1^2C_{\bar{d}LueH}^{prst}\mathcal{O}_{\bar{d}LueH}^{prst},
\\
\nonumber
\langle (C\mathcal{O})_{\bar{d}LueH}\rangle^{(j)}\delta&=&-\xi_1 y_ey_Hg_1^2C_{\bar{d}LueH}^{prst}\mathcal{O}_{\bar{d}LueH}^{prst},
\\
\nonumber
\langle (C\mathcal{O})_{\bar{d}LueH}\rangle^{(k)}\delta&=&-\frac{1}{2}(Y_d)_{pv}(Y_e^\dagger)_{ws}C_{\bar{d}LueH}^{vtrw}\mathcal{O}_{\bar{d}LueH}^{prst},
\\
\nonumber
\langle (C\mathcal{O})_{\bar{d}LueH}\rangle^{(l)}\delta&=&(Y_u^\dagger)_{vs}(Y_e^\dagger)_{wt}C_{\bar{d}LueH}^{prvw}\Big(\mathcal{O}_{\bar{d}LQLH1}^{prst}-\mathcal{O}_{\bar{d}LQLH2}^{prst}\Big),
\\
\nonumber
\langle (C\mathcal{O})_{\bar{d}LueH}\rangle^{(m)}\delta&=&-\frac{1}{2}(Y_e)_{vt}(Y_e^\dagger)_{wr}C_{\bar{d}LueH}^{pvsw}\mathcal{O}_{\bar{d}LueH}^{prst},
\\
\nonumber
\langle (C\mathcal{O})_{\bar{d}LueH}\rangle^{(n)}\delta&=&-(Y_d^\dagger Y_d)_{pv}C_{\bar{d}LueH}^{vrst}\mathcal{O}_{\bar{d}LueH}^{prst},
\\
\nonumber
\langle (C\mathcal{O})_{\bar{d}LueH}\rangle^{(o)}\delta&=&-(Y_u^\dagger Y_u)_{vs}C_{\bar{d}LueH}^{prvt}\mathcal{O}_{\bar{d}LueH}^{prst},
\\
\nonumber
\langle (C\mathcal{O})_{\bar{d}LueH}\rangle^{(p)}\delta&=&(Y_eY_e^\dagger)_{vr}C_{\bar{d}LueH}^{pvst}\mathcal{O}_{\bar{d}LueH}^{prst},
\\
\nonumber
\langle (C\mathcal{O})_{\bar{d}LueH}\rangle^{(q)}\delta&=&-\frac{1}{2}(Y_e^\dagger Y_e)_{vt}C_{\bar{d}LueH}^{prsv}\mathcal{O}_{\bar{d}LueH}^{prst},
\\
\langle (C\mathcal{O})_{\bar{d}LueH}\rangle^{(r)}\delta&=&3(Y_u^\dagger Y_d)_{vw}C_{\bar{d}LueH}^{wpvr}\mathcal{O}_{LeHD}^{pr},
\end{eqnarray}
where $\xi_{1,2,3}$ are the gauge parameters for the SM gauge group and $y_{L,e,Q,u,d,H}$ are hypercharges. Including the term due to wavefunction renormalization
\begin{eqnarray}
\nonumber
\langle (C\mathcal{O})_{\bar{d}LueH}\rangle^{(s)}\delta&=&-\Big(\frac{8}{3}\xi_3g_3^2
+\frac{3}{2}\xi_2g_2^2+\frac{37}{18}\xi_1g_1^2-\frac{9}{4}g_2^2-\frac{3}{4}g_1^2+W_H\Big)C^{prst}_{\bar{d}LueH}\mathcal{O}^{prst}_{\bar{d}LueH}
\\
&&-\Big(   (Y_d^\dagger Y_d)_{pv}C^{vrst}_{\bar{d}LueH}+\frac{1}{2}(Y_eY_e^\dagger )_{vr}C^{pvst}_{\bar{d}LueH} +(Y_u^\dagger Y_u )_{vs}C^{prvt}_{\bar{d}LueH} +(Y_e^\dagger Y_e)_{vt}C^{prsv}_{\bar{d}LueH} \Big)\mathcal{O}^{prst}_{\bar{d}LueH},
\end{eqnarray}
the complete one-loop correction is
\begin{eqnarray}
\langle (C\mathcal{O})_{\bar{d}LueH}\rangle \delta = \sum_{\alpha=a}^r\langle (C\mathcal{O})_{\bar{d}LueH}\rangle^{(\alpha)}\delta+\frac{1}{2}\langle (C\mathcal{O})_{\bar{d}LueH}\rangle^{(s)}\delta.
\end{eqnarray}
Plugging in the values of hypercharges returns the final answer shown in eq.~\eqref{examplesum}. As cross checks of our calculation we note that the final answer does not depend on gauge parameters, is consistent with perturbative power counting~\cite{Liao:2017amb}, and conforms to nonrenormalization theorem~\cite{Cheung:2015aba} when nonholomorphic Yukawa couplings are discarded.

\end{appendices}


\begin{thebibliography}{100}

\bibitem{Weinberg:1979sa}
  S.~Weinberg,
  Phys.\ Rev.\ Lett.\  {\bf 43}, 1566 (1979).

\bibitem{Buchmuller:1985jz}
  W.~Buchmuller and D.~Wyler,
  Nucl.\ Phys.\ B {\bf 268}, 621 (1986).

\bibitem{Grzadkowski:2010es}
  B.~Grzadkowski, M.~Iskrzynski, M.~Misiak and J.~Rosiek,
  JHEP {\bf 1010}, 085 (2010)
  [arXiv:1008.4884 [hep-ph]].

\bibitem{Lehman:2014jma}
  L.~Lehman,
  Phys.\ Rev.\ D {\bf 90}, 125023 (2014)
  [arXiv:1410.4193 [hep-ph]].

\bibitem{Liao:2016hru}
  Y.~Liao and X.~D.~Ma,
  JHEP {\bf 1611}, 043 (2016)
  [arXiv:1607.07309 [hep-ph]].

\bibitem{Lehman:2015via}
  L.~Lehman and A.~Martin,
  Phys.\ Rev.\ D {\bf 91}, 105014 (2015)
  [arXiv:1503.07537 [hep-ph]].

\bibitem{Henning:2015daa}
  B.~Henning, X.~Lu, T.~Melia and H.~Murayama,
  Commun.\ Math.\ Phys.\  {\bf 347}, 363 (2016)
  [arXiv:1507.07240 [hep-th]].

\bibitem{Lehman:2015coa}
  L.~Lehman and A.~Martin,
  JHEP {\bf 1602}, 081 (2016)
  [arXiv:1510.00372 [hep-ph]].

\bibitem{Henning:2015alf}
  B.~Henning, X.~Lu, T.~Melia and H.~Murayama,
  JHEP {\bf 1708}, 016 (2017)
  [arXiv:1512.03433 [hep-ph]].

\bibitem{Henning:2017fpj}
  B.~Henning, X.~Lu, T.~Melia and H.~Murayama,
  JHEP {\bf 1710}, 199 (2017)
  [arXiv:1706.08520 [hep-th]].

\bibitem{Aparici:2009fh}
  A.~Aparici, K.~Kim, A.~Santamaria and J.~Wudka,
  Phys.\ Rev.\ D {\bf 80}, 013010 (2009)
  [arXiv:0904.3244 [hep-ph]].

\bibitem{delAguila:2008ir}
  F.~del Aguila, S.~Bar-Shalom, A.~Soni and J.~Wudka,
  Phys.\ Lett.\ B {\bf 670}, 399 (2009)
  [arXiv:0806.0876 [hep-ph]].

\bibitem{Bhattacharya:2015vja}
  S.~Bhattacharya and J.~Wudka,
  Phys.\ Rev.\ D {\bf 94}, 055022 (2016)
  [arXiv:1505.05264 [hep-ph]].

\bibitem{Liao:2016qyd}
  Y.~Liao and X.~D.~Ma,
  Phys.\ Rev.\ D {\bf 96}, 015012 (2017)
  [arXiv:1612.04527 [hep-ph]].

\bibitem{Antusch:2001ck}
  S.~Antusch, M.~Drees, J.~Kersten, M.~Lindner and M.~Ratz,
  Phys.\ Lett.\ B {\bf 519}, 238 (2001)
  [hep-ph/0108005].

\bibitem{Grojean:2013kd}
  C.~Grojean, E.~E.~Jenkins, A.~V.~Manohar and M.~Trott,
  JHEP {\bf 1304}, 016 (2013)
  [arXiv:1301.2588 [hep-ph]].

\bibitem{Elias-Miro:2013gya}
  J.~Elias-Miro, J.~R.~Espinosa, E.~Masso and A.~Pomarol,
  JHEP {\bf 1308}, 033 (2013)
  [arXiv:1302.5661 [hep-ph]].

\bibitem{Elias-Miro:2013mua}
  J.~Elias-Miro, J.~R.~Espinosa, E.~Masso and A.~Pomarol,
  JHEP {\bf 1311}, 066 (2013)
  [arXiv:1308.1879 [hep-ph]].

\bibitem{Jenkins:2013zja}
  E.~E.~Jenkins, A.~V.~Manohar and M.~Trott,
  JHEP {\bf 1310}, 087 (2013)
  [arXiv:1308.2627 [hep-ph]].

\bibitem{Jenkins:2013wua}
  E.~E.~Jenkins, A.~V.~Manohar and M.~Trott,
  JHEP {\bf 1401}, 035 (2014)
  [arXiv:1310.4838 [hep-ph]].

\bibitem{Alonso:2013hga}
  R.~Alonso, E.~E.~Jenkins, A.~V.~Manohar and M.~Trott,
  JHEP {\bf 1404}, 159 (2014)
  [arXiv:1312.2014 [hep-ph]].

\bibitem{Alonso:2014zka}
  R.~Alonso, H.~M.~Chang, E.~E.~Jenkins, A.~V.~Manohar and B.~Shotwell,
  Phys.\ Lett.\ B {\bf 734}, 302 (2014)
  [arXiv:1405.0486 [hep-ph]].

\bibitem{Alonso:2014rga}
  R.~Alonso, E.~E.~Jenkins and A.~V.~Manohar,
  Phys.\ Lett.\ B {\bf 739}, 95 (2014)
  [arXiv:1409.0868 [hep-ph]].

\bibitem{Cheung:2015aba}
  C.~Cheung and C.~H.~Shen,
  Phys.\ Rev.\ Lett.\  {\bf 115}, 071601 (2015)
  [arXiv:1505.01844 [hep-ph]].

\bibitem{Jenkins:2013sda}
  E.~E.~Jenkins, A.~V.~Manohar and M.~Trott,
  Phys.\ Lett.\ B {\bf 726}, 697 (2013)
  [arXiv:1309.0819 [hep-ph]].

\bibitem{Liao:2017amb}
  Y.~Liao and X.~D.~Ma,
  Commun.\ Theor.\ Phys.\  {\bf 69}, 285 (2018)
  [arXiv:1701.08019 [hep-ph]].

\bibitem{Arzt:1993gz}
  C.~Arzt,
  Phys.\ Lett.\ B {\bf 342}, 189 (1995)
  [hep-ph/9304230].

\bibitem{Cirigliano:2017djv}
  V.~Cirigliano, W.~Dekens, J.~de Vries, M.~L.~Graesser and E.~Mereghetti,
  JHEP {\bf 1712}, 082 (2017)
  [arXiv:1708.09390 [hep-ph]].
\bibitem{Cirigliano:2018yza}
V.~Cirigliano, W.~Dekens, J.~de Vries, M.~L.~Graesser and E.~Mereghetti,
JHEP {\bf 1812}, 097 (2018)
[arXiv:1806.02780 [hep-ph]].


\bibitem{CERNNA48/2:2016tdo}
  J.~R.~Batley {\it et al.} [NA48/2 Collaboration],
  Phys.\ Lett.\ B {\bf 769}, 67 (2017)
  [arXiv:1612.04723 [hep-ex]].

\bibitem{Rodejohann:2011mu}
  W.~Rodejohann,
  Int.\ J.\ Mod.\ Phys.\ E {\bf 20}, 1833 (2011)
  [arXiv:1106.1334 [hep-ph]].

\bibitem{KamLAND-Zen:2016pfg}
  A.~Gando {\it et al.} [KamLAND-Zen Collaboration],
  Phys.\ Rev.\ Lett.\  {\bf 117}, 082503 (2016)
  Addendum: [Phys.\ Rev.\ Lett.\  {\bf 117}, 109903 (2016)]
  [arXiv:1605.02889 [hep-ex]].

\bibitem{Agostini:2018tnm}
  M.~Agostini {\it et al.} [GERDA Collaboration],
  Phys.\ Rev.\ Lett.\  {\bf 120}, 132503 (2018)
  [arXiv:1803.11100 [nucl-ex]].

\bibitem{Kharusi:2018eqi}
  S.~A.~Kharusi {\it et al.} [nEXO Collaboration],
  arXiv:1805.11142 [physics.ins-det].

\bibitem{Pas:2000vn}
  H.~Pas, M.~Hirsch, H.~V.~Klapdor-Kleingrothaus and S.~G.~Kovalenko,
  Phys.\ Lett.\ B {\bf 498}, 35 (2001)
  [hep-ph/0008182].

\bibitem{Deppisch:2012nb}
  F.~F.~Deppisch, M.~Hirsch and H.~Pas,
  J.\ Phys.\ G {\bf 39}, 124007 (2012)
  [arXiv:1208.0727 [hep-ph]].

\bibitem{Horoi:2017gmj}
  M.~Horoi and A.~Neacsu,
  arXiv:1706.05391 [hep-ph].

\bibitem{Tanabashi:2018pdg}
  M. Tanabashi et al. (Particle Data Group), Phys. Rev. D 98, 030001 (2018).

\bibitem{Mihaila:2012pz}
  L.~N.~Mihaila, J.~Salomon and M.~Steinhauser,
  Phys.\ Rev.\ D {\bf 86}, 096008 (2012)
  [arXiv:1208.3357 [hep-ph]].
\bibitem{Nicholson:2018mwc}
A.~Nicholson {\it et al.},
Phys.\ Rev.\ Lett.\  {\bf 121}, 172501 (2018)
[arXiv:1805.02634 [nucl-th]].

\bibitem{Feng:2018pdq}
X.~Feng, L.~C.~Jin, X.~Y.~Tuo and S.~C.~Xia,
Phys.\ Rev.\ Lett.\  {\bf 122}, 022001 (2019)
[arXiv:1809.10511 [hep-lat]].


\end{thebibliography}
\end{document}